\documentclass{aa}
\usepackage{graphicx, psfig, epsf, epsfig}

\newbox\grsign \setbox\grsign=\hbox{$>$}
\newdimen\grdimen \grdimen=\ht\grsign
\newbox\laxbox \newbox\gaxbox
\setbox\gaxbox=\hbox{\raise.5ex\hbox{$>$}\llap
        {\lower.5ex\hbox{$\sim$}}}\ht1=\grdimen\dp1=0pt
\setbox\laxbox=\hbox{\raise.5ex\hbox{$<$}\llap
        {\lower.5ex\hbox{$\sim$}}}\ht2=\grdimen\dp2=0pt
\newcommand{\simlt}{\mathrel{\copy\laxbox}}
\newcommand{\simgt}{\mathrel{\copy\gaxbox}} 

\begin{document}
   \title{X-ray sources in the starburst spiral galaxy M83}

   \subtitle{Nuclear region and discrete source population}

   \author{Roberto Soria \inst{1,2} \and
          Kinwah Wu \inst{1,2}
          }

   \offprints{R.~Soria ({\tt rs1@mssl.ucl.ac.uk})}

   \institute{Mullard Space Science Laboratory, 
          University College London, Holmbury St Mary, 
          Surrey RH5 6NT, UK  \\
              emails: {\tt{rs1@mssl.ucl.ac.uk}}, {\tt{kw@mssl.ucl.ac.uk}}  
         \and School of Physics A28, University of Sydney, 
           NSW 2006, Australia} 


   \date{Received ; accepted }

   \abstract{  
      {\em Chandra} has resolved the starburst nuclear region 
        of the face-on grand-design spiral M83. 
      Eighty-one point sources are detected (above 3.5-$\sigma$) 
        in the ACIS S3 image, 
        and 15 of them are within the inner 16\arcsec~region of the galaxy.  
      A point source with $L_{\rm x} \approx 3 \times 10^{38}$~erg~s$^{-1}$ 
        in the 0.3--8.0 keV band is found to coincide 
        with the infra-red nuclear photometric peak,  
        one of the two dynamical nuclei of the galaxy.  
        No point-like sources are resolved (at a 2.5-$\sigma$ level) 
        at the centre of symmetry of the outer optical isophote ellipses, 
        suspected to be another dynamical nucleus.   
      About 50\% of the total emission in the nuclear region  
        is unresolved; of this, about 70\% can be attributed 
	to hot thermal plasma, and the rest is probably due to unresolved 
        point sources (eg, faint X-ray binaries). 
      The azimuthally-averaged radial distribution 
	of the unresolved emission 
        has a King-like profile, with no central cusp.  
      Strong emission lines are seen in the spectrum of 
	the optically thin plasma component.   
      The high abundances of C, Ne, Mg, Si and S with respect to Fe 
        suggest 
        that the interstellar medium in the nucleus is enriched and heated  
        by type-II supernova explosions and winds from massive stars. 
      The cumulative luminosity distribution of the discrete X-ray sources 
        is neither a single nor a broken power law.     
      Separating the sources in the nuclear region 
        (within a distance of 60\arcsec~from the X-ray centre) 
	from the rest reveals that the two groups 
	have different luminosity distributions.    
      The log~N($>$S) -- log S curve of the sources 
in the inner region (nucleus and stellar bar) is a single power law, 
which we interpret as due to continuous, ongoing star formation. 
Outside the central region, there is 
a smaller fraction of sources brighter than the Eddington limit 
for an accreting neutron star.
   \keywords{  
      Galaxies: individual: M83 (=NGC~5236) --  
      Galaxies: nuclei --  
      Galaxies: spiral -- 
      Galaxies: starburst --         
      X-rays: binaries --  
      X-rays: galaxies }
}

\authorrunning{R. Soria \& K. Wu}
   \maketitle
%

\section{Introduction}  

M83 (NGC~5236) is a grand-design, barred spiral galaxy 
   (Hubble type SAB(s)c) with a starburst nucleus.  
Distance estimates are still very uncertain.
A value of 3.7~Mpc was obtained 
   by de Vaucouleurs et al.\ (1991).    
This places the galaxy in the Centaurus A group,  
  whose members have a large spread in morphology and high velocities,  
  indicating that the group is not virialised 
  and tidal interactions and merging are frequent 
  (de Vaucoulers 1979; C{\^ o}t{\' e} et al.\ 1997).  
A distance of 8.9~Mpc was instead given in Sandage \& Tamman (1987).  

Infra-red (IR) observations 
  (Gallais et al.\ 1991; Elmegreen, Chromey \& Warren 1998; 
  Thatte, Tecza \& Genzel 2000) have shown 
  that the nuclear region of M83 has a complex structure.     
From measurements of line-of-sight stellar velocities  
  in the inner galactic region, 
  two dynamical centres are inferred.   
The first centre is identified 
  with a strong point-like optical/IR source. 
The second centre, located $1\farcs5$ to the South 
  and $3\farcs0$ to the West of the IR peak, 
  is not associated with any bright source, 
  but is approximately coincident 
  with the centre of symmetry of the outer isophote ellipses.   
The stellar velocity dispersion  
  implies that each dynamical centre contains 
  an enclosed mass of $\approx 1.3 \times 10^{7}$ M$_{\odot}$ 
  (Thatte et al.\  2000).   

The $J-K$ images of the nuclear region (Elmegreen et al.\ 1998)
  show two circumnuclear dust rings. 
The inner one has a radius of $2\farcs8$ 
  and is centred on the IR nuclear peak; 
  the outer one has a radius of $8\farcs6$ and is centred $2\farcs5$ 
  South-West of the IR nucleus. 
The two rings are connected by a mini bar, 
  oriented almost perpendicularly to the main galactic bar. 
Starburst activity is concentrated 
  in a semi-circular annulus located 
  $\approx 7$\arcsec~South-West of the IR nucleus, 
  just inside the South-West half of the outer dust ring 
  (Elmegreen et al.\ 1998; Harris et al.\ 2001).

M83 was observed in the X-ray bands by {\em Einstein} 
  in 1979--1981 (Trinchieri, Fabbiano \& Palumbo 1985), 
  by {\em ROSAT} in 1992--1994, and 
  by {\em ASCA} in 1994 (Okada, Mitsuda \& Dotani 1997).   
Thirteen point sources were found in the  
  {\em ROSAT} PSPC image (Ehle et al.\ 1998)  
  and 37 in the {\em ROSAT} HRI image.  
After removing probable foreground stars and background AGN,   
  there are 21 sources within the D$_{25}$ ellipse 
  believed to belong to the galaxy (Immler et al.\ 1999). 
The nuclear region,  
  which encloses approximately 25\% of the total X-ray luminosity  
  in the 0.1--2.4~keV band, 
  was unresolved in the {\em ROSAT} images.    

M83 was observed by {\em Chandra} in 2000 April, 
  and the data became available to the public in mid-2001. 
In this paper we report the results 
  of our analysis of the {\em Chandra} archival data.    
We discuss the source population in the galaxy 
  and the properties of discrete sources and unresolved emission 
  in the nuclear region.


\begin{figure} 
\vspace*{0.25cm} 
\psfig{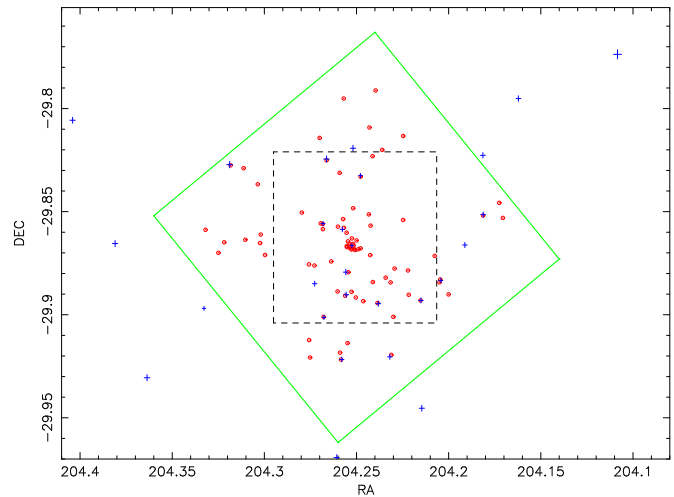}
\caption{
  Spatial distribution of the {\em Chandra} sources (red circles)   
      detected in the S3 chip (green square), 
      and of the {\em ROSAT} HRI sources 
      (blue crosses with sizes denoting the positional uncertainty). 
The central region delimited by a dashed box is shown in 
greater detail in Figure~2. }
\end{figure}


\section{Data analysis}    

The {\em Chandra} observation was carried out on 2000 April 29 
  (Observation ID: 793; PI: G.\ Rieke), 
  with the ACIS-S3 chip at the focus 
  and a total exposure time of 50.978~ks.  
After screening out observational intervals corresponding to background 
flares, we retained a good time interval of 49.497~ks.  
The CXC {\small{CIAO}} software (version 2.2) was used in the analysis.  

Discrete X-ray sources were identified 
  using the {\small{CIAO}} source-finding routine {\it celldetect}. 
The percentage of the PSF energy encircled by the detect cells 
was chosen to be 80\% at 1.5~keV.  
We did not select a higher percentage 
  in order to avoid confusion of close-by sources 
  in the nuclear region and of a few off-centre close pairs  
  (eg, the {\em ROSAT} HRI source H17, Immler et al.\ 1999).   
We compared the results obtained with {\it celldetect} 
  with those obtained by using {\it wavdetect}, and  
  we found no significant differences  
  for sources away from the galactic centre,   
  for a signal-to-noise ratio $> 3.5$. 
The routine {\it celldetect}, however, 
  seems more successful at resolving the nuclear sources.   
Exposure maps at energies of 1.0, 1.7 and 3.0~keV 
  were first calculated, and then used in the routine {\it dmextract} 
  to correct the net count rates of the sources.

\section{Global properties of the sources}     

A total of 81 point sources 
  are detected in the S3 chip at a 3.5-$\sigma$ level 
in the 0.3--8.0~keV band. 
Their positions, with an uncertainty of 0\farcs5, 
  their total photon counts and their counts in three separate energy bands 
(0.3--1.0 keV, 1.0--2.0 keV and 2.0--8.0 keV) are listed in Table A.1.    
The positions of the {\em ROSAT} sources  
  given in Table 3 of Immler et al.\ (1999) 
  have a systematic offset 
  $\Delta ({\rm R.A.}) \simeq -0^{\rm s}.5$ and $\Delta ({\rm Dec.}) \simeq 3\farcs5$
  with respect to the {\em Chandra} S3 source positions. 
Therefore, we shifted the positions of the {\em ROSAT} sources 
by the same amount (without applying any rotation),   
  before comparing them with the {\em Chandra} sources, 
  as shown in the spatial distribution plot (Figure~1).


\begin{figure} 
\vspace*{0.25cm} 
\psfig{figure=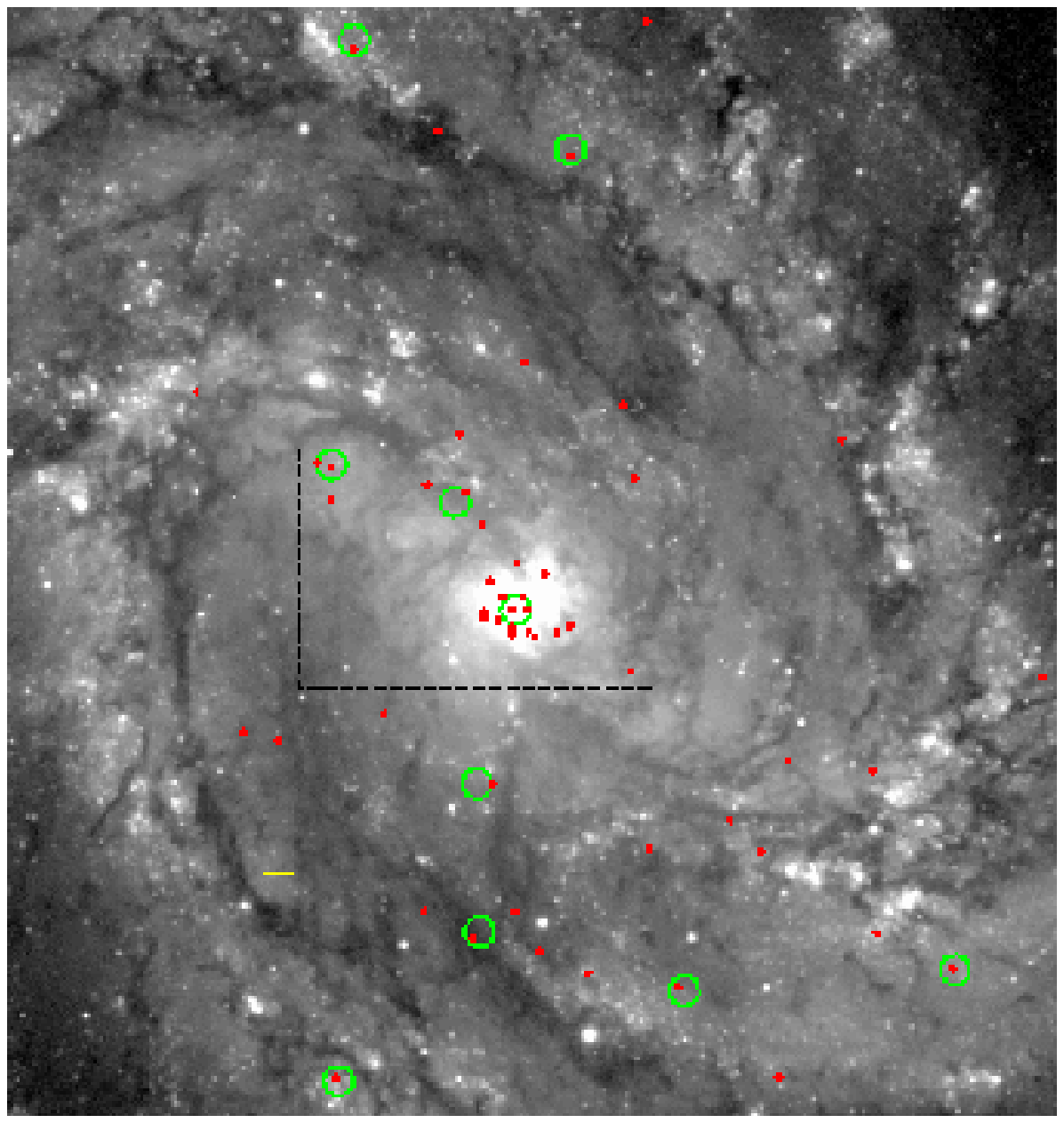,width=8.8cm} 
\psfig{figure=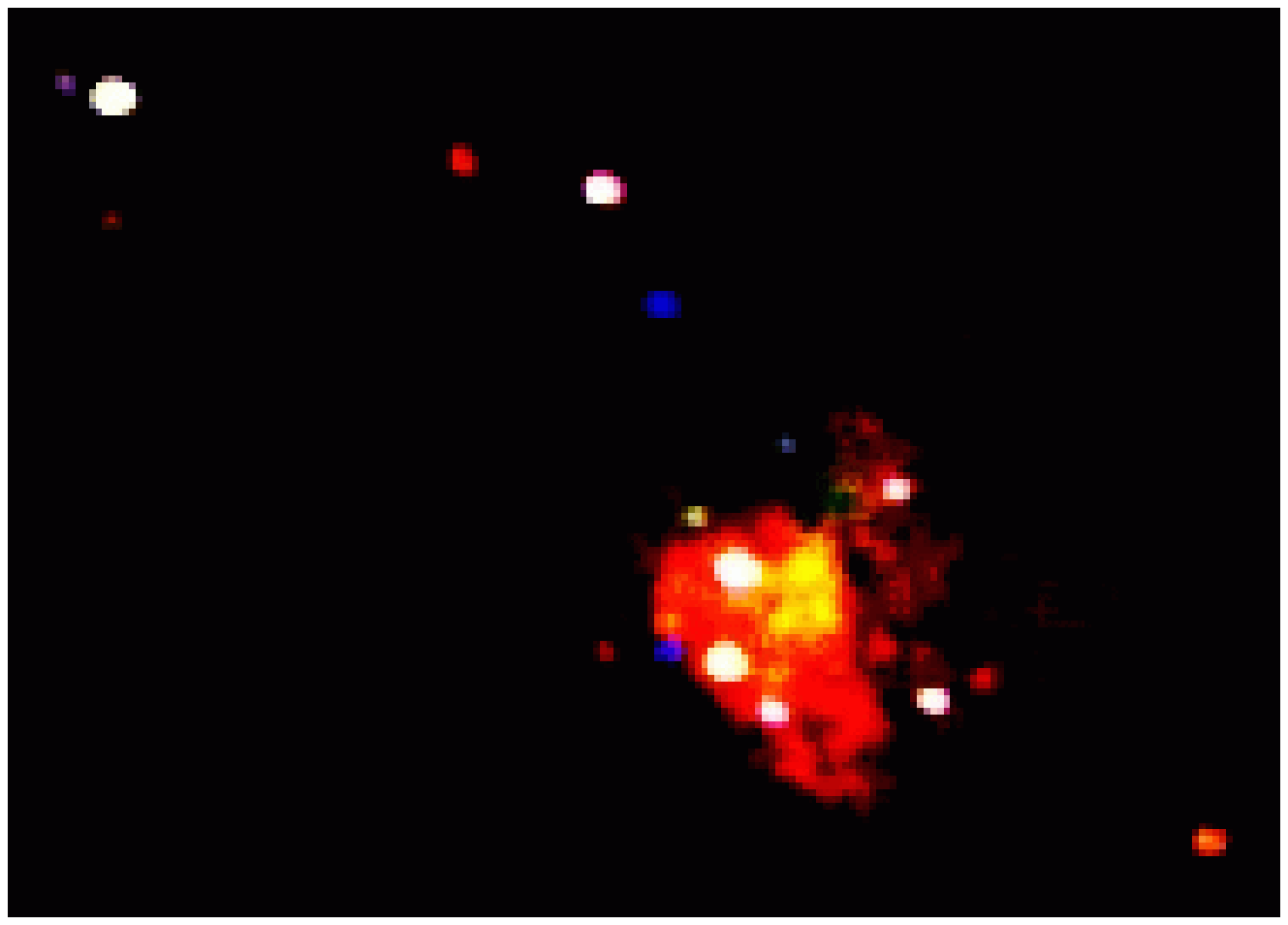,width=8.8cm}
\caption{
  Top panel:   
  positions of the discrete X-ray sources overplotted 
      on a Very Large Telescope (VLT) $B$ image 
      (inner $4\farcm8 \times 5\farcm0$ region of M83,    
      including the main bar). 
  The image was obtained from the ESO public archive; 
it was taken on 1999 March 11 with the FORS1 camera 
      at Unit Telescope 1.    
  The {\em Chandra} sources are represented by red circles   
      (radius $= 1\arcsec$),    
      {\em ROSAT} sources detected by {\em Chandra} 
      by green circles (radius $= 4\arcsec$), 
      and a {\em ROSAT} source not detected by {\em Chandra}  
      by a yellow box (size $= 8\arcsec \times 8\arcsec$).  
  Bottom panel:
  ``true-colour'' image of the nuclear region 
      (delimited by the dashed box in the VLT $B$ image), 
smoothed using a $1\farcs5 \times 1\farcs5$ boxcar 
kernel.  Red, green and blue correspond to emission in the 0.3--1.0~keV, 
      1.0--2.0~keV and 2.0--8.0~keV bands respectively.
In both images, North is up and East is left.}
\label{fig:2}
\end{figure}


Fifteen of the 18 {\em ROSAT} HRI sources   
  within the field of view of the {\em Chandra} S3 chip are detected again. 
{\em Chandra} has resolved at least 15 sources  
  in the nuclear region, 
  which was not resolved in {\em ROSAT} HRI and PSPC images. 
It has also resolved two off-centre close pairs:  
  the {\em ROSAT} HRI sources H12 and H27.   
The {\em ROSAT} HRI sources H11, H18 and H28 were undetected 
at a $2.5$-$\sigma$ level.   
If we assume an absorbing column density 
  $n_{\rm H} = 4 \times 10^{20}$ cm$^{-2}$
  (the Galactic foreground absorption, 
from Schlegel, Finkbeiner \& Davis 1998 and Predehl \& Schmitt 1995) 
  and a power-law spectrum with photon index $\Gamma = 1.5$ for all the sources,
  the detection threshold of $\simeq 15$ net counts  
  places an upper limit of  
  $f_{{\rm x}} \simeq 2.0 \times 10^{-15}$~erg cm$^{-2}$~s$^{-1}$  
  for the observed flux in the 0.3--8.0~keV band. 
Those three transients were therefore at least an order of magnitude 
  fainter than they were during the {\em ROSAT} observations.    

Most of the sources are concentrated in the nuclear region (Figures~1, 2). 
Comparing the position of the {\em Chandra} S3 sources 
with a VLT $B$ image  
  shows that the off-centre sources tend to associate 
  with the spiral arms (Figure 2, top panel).  
The sources have a large spread in the hardness of their X-ray emission.  
  A ``true-colour'' X-ray image of the nuclear region is shown 
  in Figure 2, bottom panel. 

Separating the sources inside and outside a circular region 
  of radius 60\arcsec~from the geometric centre of the X-ray emission   
  reveals that the two groups have different luminosity distributions 
in the 0.3--8.0~keV band.
(A linear separation of 60\arcsec~corresponds to 1.1~kpc 
  for a distance of 3.7~Mpc, 
  and is roughly half of the total length of the major galactic bar.)
The cumulative log~N($>$S) -- log~S distribution 
  (where S are the photon counts) 
  of the sources outside the circular inner region  
  is neither a single nor a broken power law (Figure 3).       
It shows a kink at S $\approx 250$~cts;
the slope of the curve above this feature is $-1.3$, while 
it is -0.6 at the faint end.
The log~N($>$S) -- log~S curve of the inner sources  
  can instead be described as a single power law,     
with a slope of $-0.8$.
At an assumed distance of 3.7~Mpc, 
  100~counts ($\simeq 2.0 \times 10^{-3}$~cts~s$^{-1}$) 
  correspond to an unabsorbed source luminosity 
  $L_{\rm x} = 2.3 \times 10^{37}$~erg~s$^{-1}$ in the 0.3--8.0~keV band.  
The kink in the log~N($>$S) -- log~S curve 
  of the sources outside the 60\arcsec~circle 
  is therefore located at 
$L_{\rm x} \approx 6 \times 10^{37}$~erg~s$^{-1}$ (0.3--8.0~keV band).


\begin{figure} 
\vspace*{0.25cm}
\psfig{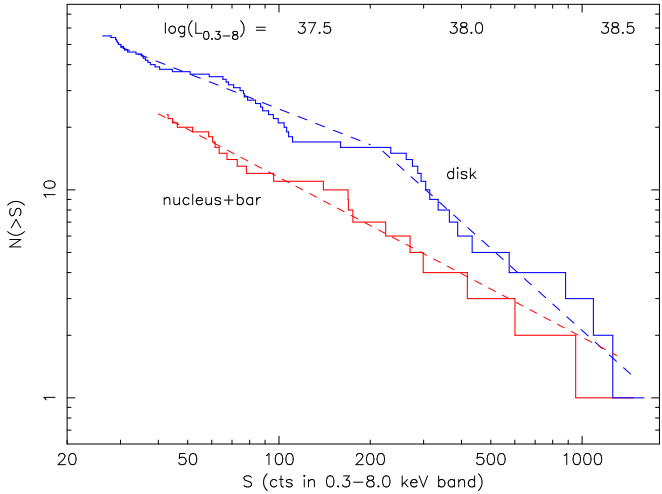}
\caption{
The cumulative luminosity distributions of sources found 
in the 0.3--8.0~keV band is different for the population  
inside the 60\arcsec~inner circle (which includes the starburst 
nuclear region and the main bar) and for the population outside of it (red 
and blue curves respectively). We interpret the single power-law 
distribution of the nuclear sources as evidence of continuous, ongoing 
star formation. The relative scarcity of bright sources in the disk 
population suggests that star formation is less active there. 
See Section 3 for the conversion from counts to luminosity.}
\label{fig:lumf}
\end{figure} 
  

Some of the sources are likely to be background AGN. In order to  
investigate the effect of this correction, we have constructed  
the log~N($>$S) -- log~S curves in a soft (0.5--2.0 keV) 
and hard (2.0--10.0 keV) band separately, and we have used 
the luminosity functions of faint background sources in the two bands, 
deduced from the Deep Field South survey (Giacconi et al.\ 2001).   
We find that about 15\% of the sources are background AGN; 
the expected number in the inner 60\arcsec~circle  
   is smaller than one. The kink in the log~N($>$S) -- log~S curve 
   for the outer sources and the values of the slope at both ends 
are unaffected by the background subtraction.
Because of the uncertainty in the distant AGN counts 
   in different sky regions, 
   the background correction itself is also uncertain; 
however, it is unlikely that anisotropic effects 
   will significantly alter the results that we have presented above.


\begin{figure*}
\vbox{
\begin{tabular}{lr}
\psfig{figure=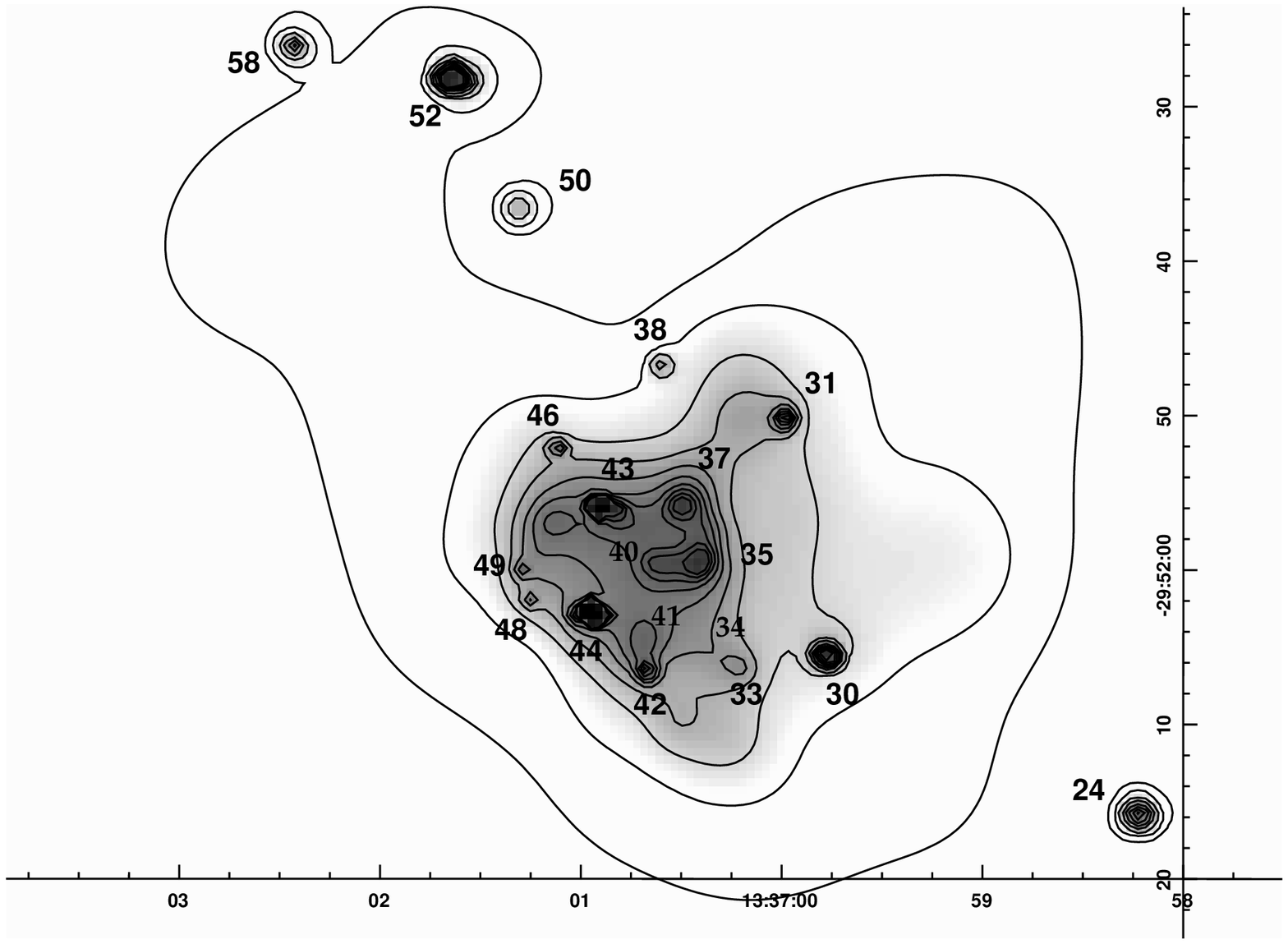,width=8.6cm} &
\psfig{figure=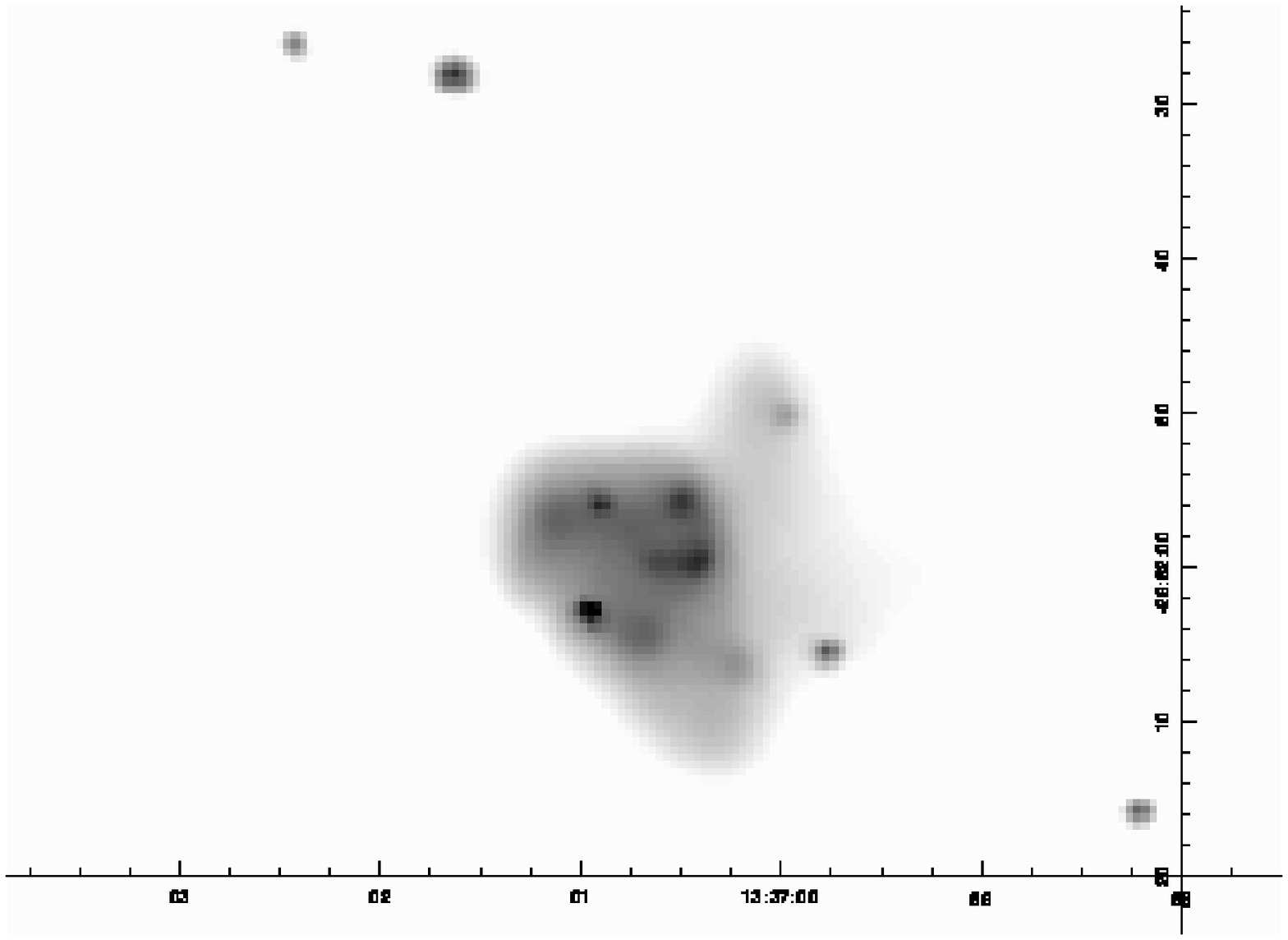,width=8.6cm} \\
\psfig{figure=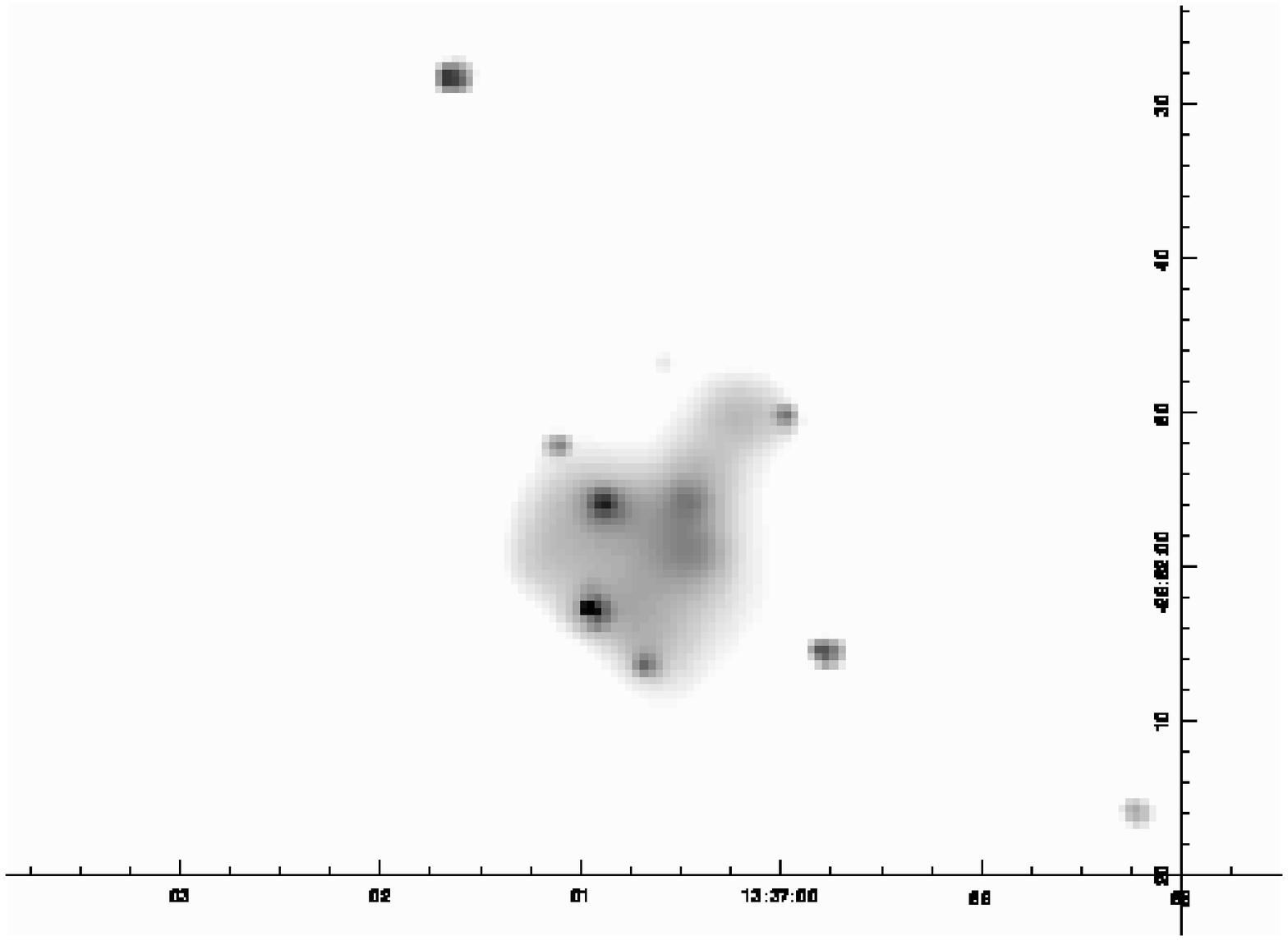,width=8.6cm} &
\psfig{figure=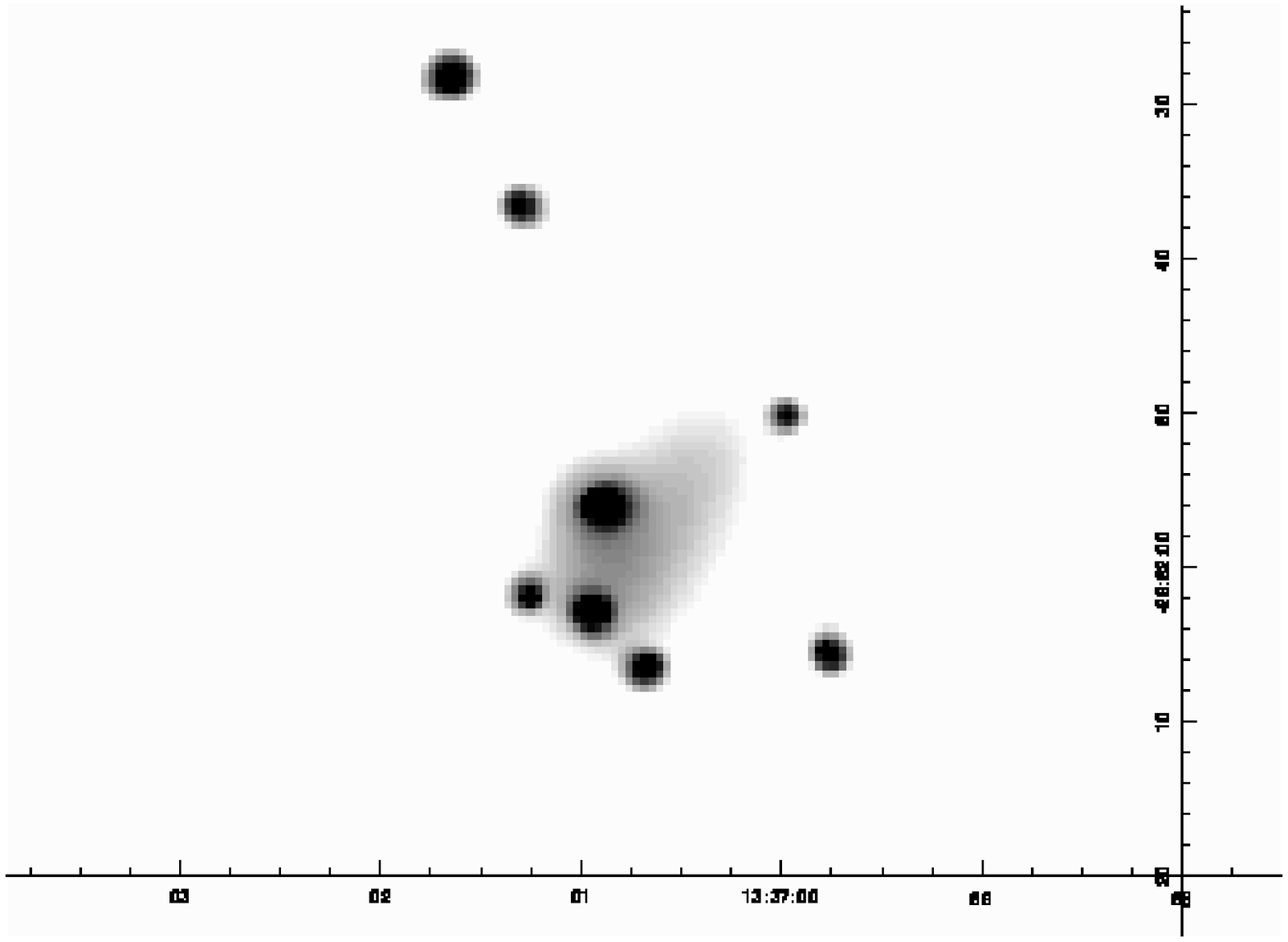,width=8.6cm} \\
\end{tabular}
}
\caption{Top left:  
  image of the nuclear region of M83 in the 0.3--8.0~keV band, 
with the contours of the X-ray emission overplotted in square-root scale;   
the numbers identify the sources listed in Table A.1.
Top right: image in the 0.3--1.0~keV band, showing extended emission 
to the West of the star-forming arc. Bottom left: image in the 1.0--2.0~keV band. 
Bottom right: image in the 2.0--8.0~keV band.
All images have been adaptively smoothed ({\small{CIAO}} task {\it csmooth}, 
with minimum signal-to-noise ratio $= 4$).}
\label{fig:test}
\end{figure*}


The number density of sources is about 7~kpc$^{-2}$ 
   in the inner 60\arcsec~circular region, 
   and is only 0.7~kpc$^{-2}$ elsewhere in the S3 field, 
without any background subtraction.  
After the background sources are subtracted, 
   the number density of the sources 
   detected in the 0.3--8.0~keV band  
   is 6~kpc$^{-2}$ inside the circle  
   and 0.6~kpc$^{-2}$ outside.  
The source concentration is thus at least 10 times higher 
   in the inner 60\arcsec~($\approx 1$ kpc)
   than elsewhere in the galaxy.  
        
\section{Nuclear region}

The nuclear region of M83 had not been clearly resolved 
  in X-ray imaging observations  
  (see eg, the {\em ROSAT} HRI image, Immler et al.\ 1999) 
  before {\em Chandra}.    
The {\em Chandra} data now reveal that 
  M83 has a very highly structured nuclear region 
  (Figure~2, bottom panel, and Figure~4).  
Fifteen discrete sources are detected   
  within a radius of $\simeq 16\arcsec$~($\simeq 290$ pc)
  from the centre of symmetry of the outer optical isophotes.   
There are also clump-like features in the predominantly soft 
unresolved emission, 
  which may be due to faint point sources or hot gas clouds.    
The strongest X-ray emission  
  comes from an approximately circular region, with a radius of $\simeq 7\arcsec$~and 
  a geometrical centre 
at {R.A.~(2000) $=$ 13$^h$\,37$^m$\,00$^s$.8}, 
{Dec.~(2000) $=$ $-$29$^{\circ}$\,51\arcmin\,59\farcs3} 
The region is inside the outer dust ring (Elmegreen et al.\ 1998), 
  and it contains both the IR photometric nucleus (to the North-East) 
  and the star-forming ring (to the South-West). 
The centre of the circle is less than 2\arcsec~away 
  from the invisible dynamical nucleus and the geometrical 
centre of the outer dust ring.   
The soft (0.3-1.0~keV) unresolved emission 
  is more spatially extended than 
  the hard (2.0-8.0~keV) emission 
  (Figure 4).      

We removed the point sources using the extraction regions from 
{\it celldetect} (see Section 2) and extracted counts 
  from concentric annuli to construct radial brightness profiles  
  of the unresolved emission in the nuclear region.  
We have found that the brightness is approximately constant  
  in a circular region up to a radius of 7\arcsec~and 
  then declines radially with a power-law like profile 
  (Figure~5). 
Single power-law and gaussian profiles do not provide acceptable fits 
  to the azimuthally-averaged profile.   
Better fits are obtained using a King profile, 
  yielding a core radius of $6\farcs7 \pm 0\farcs5$ ($\approx 120$~pc), 
  and a power law with a slope of $-1.9 \pm 0.2$ 
  beyond the core.


\begin{figure} 
\vspace*{0.25cm}
\psfig{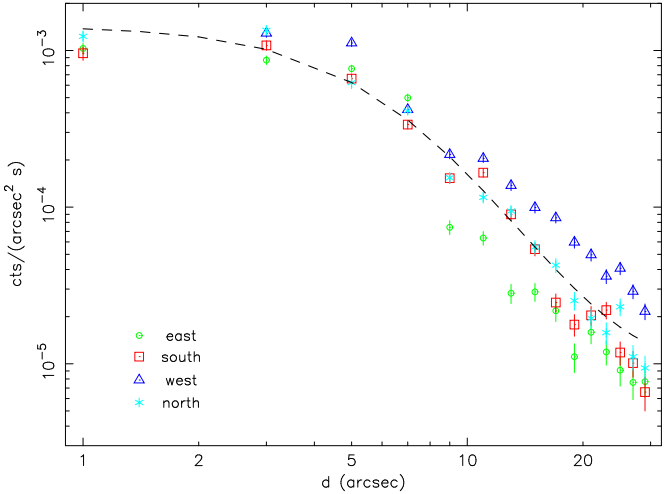}         
\caption{
   Radial profiles of the X-ray surface brightness in the nuclear region, 
     centred at {R.A.~(2000) $=$ 13$^h$\,37$^m$\,00$^s$.8}, 
{Dec.~(2000) $=$ $-$29$^{\circ}$\,51\arcmin\,59\farcs3}. 
   The profiles in the North, South, East and West quadrants 
     are represented 
     by light blue asterisks, red squares, green circles 
     and blue triangles respectively. 
   The best-fit King model (dashed line) 
     for the azimuthally-averaged radial brightness profile 
     is also shown to illustrate the relative anisotropy.  }
\label{fig:test}
\end{figure} 


The brightness distribution is not axisymmetric.  
If we divide the region into four different quadrants, 
  defined by taking the directions parallel and perpendicular 
  to the main stellar bar (which has a position angle $=51^{\circ}.7$), 
  the brightness is found to drop more steeply East of the nucleus  
  and the emission is more extended to the West (Figure~5).  
The difference in the brightness profiles of the four quadrants  
  is in fact mainly due to the extended emission region 
  westward of the star-forming arc.   
We have also examined the radial brightness profiles 
  obtained by choosing various other reference centres 
  (eg, the two dynamical nuclei inferred from IR observations),  
  but none of them appears simpler or more axisymmetric.

We extracted the spectrum of the unresolved emission inside 
the inner 16\arcsec~circle, excluding the resolved point sources, 
and we fitted it using an absorbed, single-temperature 
vmekal plus power-law model  
  ({\small{XSPEC}}, version 11.0.1, Arnaud 1996). 
Assuming solar abundances, we obtain 
the best-fit parameters listed in Table 1 (``model 1''). 
The predicted lines are not strong enough  
  to account for the data, leading to poor fit statistics 
($\chi^2_{\nu} = 1.42$, 114~dof).
Increasing the abundance of all the metals by the same constant 
factor does not improve the $\chi^2_{\nu}$. 

We then assumed a different set of abundances (Table 1, ``model 2''), 
and fitted again a vmekal plus power-law model
to the spectrum of the unresolved emission.
Our choice of higher abundances for C, Ne, Mg, Si and S 
relative to Fe is physically justified if the interstellar medium 
has been enriched by type-II supernova ejecta and winds from 
very massive, young stars (see also Section 6.2).
For this set of abundances, the model gives a total (i.e., Galactic 
foreground plus intrinsic) absorption column density
  $n_{\rm H} = (1.1^{+0.2}_{-0.3}) \times 10^{21}$ cm$^{-2}$, 
temperature $kT = (0.58^{+0.03}_{-0.02})$~keV, and power-law 
photon index $\Gamma = 2.7^{+0.3}_{-0.3}$   
  ($\chi^2_{\nu} = 0.99$, 114~dof).
In any case, the absorption column density, thermal plasma temperature 
  and power-law photon index    
  are only weakly dependent on the precise abundances. 

We left the redshift as a free parameter in our spectral fitting. 
We obtain that the emission lines are redshifted, with 
projected radial velocities $\approx 7000$ km s$^{-1}$, 
also almost independent of the choice of metal abundances. 
This is far in excess of the systemic radial velocity of M83 
($v_{\rm r} = 505$ km s$^{-1}$, Tilanus \& Allen 1993). 
A spatially resolved determination of the hot gas dynamics 
is left to further work.


\begin{figure} 
\vspace*{0.25cm} 
\psfig{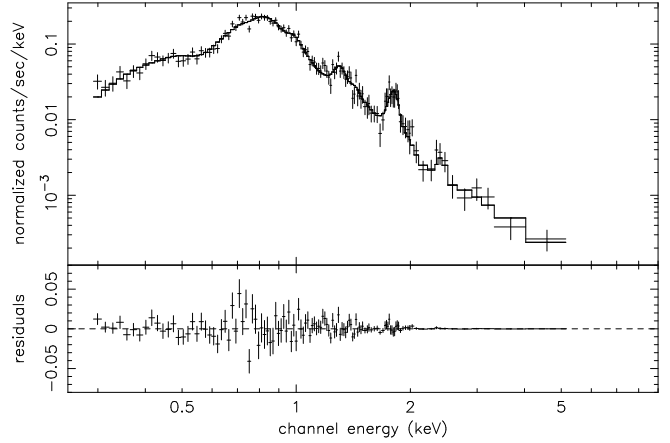}  
\caption{
The spectrum of the unresolved emission in the nuclear region (inside a 16\arcsec~circle)   
  shows very strong emission lines from 
  Mg\,{\small{XI}} (1.33--1.35~keV), Si\,{\small{XIII}} (1.84--1.87~keV) and S\,{\small{XV}} (2.43 keV).
The other major contributions to the spectrum come from 
C\,{\small{VI}} (0.37--0.44~keV), 
O\,{\small{VII}} (triplet, 0.56--0.57~keV), 
Ne\,{\small{IX}} (triplet, 0.91--0.92~keV), 
Ne\,{\small{X}} (1.02~keV), 
and the Fe\,{\small{XVII}} L-line complex (0.6--0.9~keV).
The spectrum is well fitted by an absorbed, single-temperature vemkal plus 
power-law model; the fit parameters are listed in Table~1 (``model 2''). 
The background-subtracted spectrum has been grouped 
to achieve a S/N ratio $\geq 3$ for each bin.}
\label{diffuse}
\end{figure} 


\begin{table}
\caption{
  Parameters used in the model spectral fits for  
     the unresolved emission in the inner 16\arcsec, 
     with two different sets of abundances. }
\label{mathmode} 
\centering 
\begin{tabular}{@{}lrr} 
\hline
\hline \\ 
\multicolumn{3}{c}{model: tbabs$_{\rm Gal}$~$\times$ tbabs $\times$ (power-law $+$ vmekal)}\\[5pt]  	
\hline 
\hline \\ 
  parameter &  model 1 & model 2 \\ [5pt]
\hline \\
$n_{\rm H}$~($\times 10^{21}$~cm$^{-2}$)      & $1.6^{+0.3}_{-0.3}$ &
	$1.1^{+0.2}_{-0.3}$ \\[5pt]
$\Gamma$        & $3.1^{+0.1}_{-0.2}$ & $2.7^{+0.3}_{-0.3}$ \\[5pt]
$K_{\rm pl}~(\times 10^{-5})$  & $7.6^{+1.5}_{-0.9}$  
    & $3.9^{+0.8}_{-0.9}$ \\ [5pt] 
$T$~(keV)    & $0.60^{+0.02}_{-0.03}$
	& $0.58^{+0.03}_{-0.02}$  \\[5pt]
$K_{\rm vm}~(\times 10^{-5})$   & $13.7^{+2.0}_{-0.6}$
	& $12.5^{+0.8}_{-1.2}$ \\[5pt]              
$V_{\rm r}$~(km s$^{-1}$) & $5980^{+4770}_{-1470}$& $7270^{+1900}_{-2310}$\\[5pt]
C~(fixed) & 1.0& 10.0\\[5pt]
N~(fixed)  & 1.0 & 1.0 \\[5pt]
O~(fixed)  & 1.0 & 1.0 \\[5pt]
Ne~(fixed)  & 1.0 & 2.0 \\[5pt]
Na~(fixed)  & 1.0 & 1.0 \\[5pt]
Mg~(fixed)  & 1.0 & 2.5 \\[5pt]
Al~(fixed)  & 1.0 & 1.0 \\[5pt]
Si~(fixed)  & 1.0 & 3.0 \\[5pt]
S~(fixed)  & 1.0 &  3.0\\[5pt]
Ar~(fixed)  & 1.0 & 1.0 \\[5pt]
Ca~(fixed)  & 1.0 & 1.0 \\[5pt]
Fe~(fixed)  & 1.0 & 0.9 \\[5pt]
Ni~(fixed)  & 1.0 & 1.0 \\[5pt]   
\hline \\ 
$\chi_\nu^2$~(dof)  & $1.42$~(114) & 
	$0.99$~(114)\\[5pt]
\hline 
\end{tabular}   
\end{table}  

We have also calculated the total (resolved plus unresolved) 
luminosity from the circular regions within radii 
of 7\arcsec~(this is approximately the region inside the outer dust ring) 
and 16\arcsec~from the geometric centre of the X-ray emission, 
using an absorbed, optically-thin thermal plasma  
plus power-law model. The total emitted luminosity 
in the 0.3--8.0 keV band is $\simeq 15.7 \times 10^{38}$~erg~s$^{-1}$ 
inside 7\arcsec~and $\simeq 23.8 \times 10^{38}$~erg~s$^{-1}$ 
inside 16\arcsec~(Table 2). Discrete sources contribute 
$\simeq 50\%$ of the total luminosity.

The unresolved emission is itself the sum of truly diffuse emission 
from optically thin gas, and emission from unresolved point-like 
sources (eg, faint X-ray binaries). Assuming that the latter 
contribution is responsible for the power-law component 
in the spectrum of the unresolved emission, we estimate that 
emission from truly diffuse thermal plasma contributes $\simeq 35\%$ 
of the total luminosity (Table 2).

Extrapolating the log~N($>$S) -- log~S curve for the nuclear sources 
(see Section 3) gives us another way of estimating 
the relative contribution to the 
unresolved emission of truly diffuse gas and faint point-like   
sources. 
  We obtain that unresolved point-like X-ray sources 
brighter than $10^{34}$~erg~s$^{-1}$ 
  would contribute $\approx 4.2 \times 10^{38}$~erg~s$^{-1}$ 
  to the luminosity in the inner 60\arcsec~region.   
Of the 23 discrete sources identified within the 60\arcsec~radius, 
15 are found within the inner 16\arcsec. 
Assuming that the same uniform scaling applies 
  to the emission of both resolved and unresolved stellar sources, 
  this implies that 
  unresolved point-like sources inside 16\arcsec~would have a total luminosity 
  of $\approx 2.7 \times 10^{38}$ erg s$^{-1}$. 

Another possible contribution to the unresolved emission detected 
inside the inner 16\arcsec~comes from photons emitted by the resolved sources 
but falling outside the extraction regions, in the wings 
of the PSF. Taking into account the partial spatial overlapping of the 
detection cells, we estimate that this contribution is $\simlt 1.5 \times 
10^{38}$~erg~s$^{-1}$. 

Thus, the combined contribution of faint X-ray sources and emission in 
the wings of the PSF can account for the power-law component inferred 
from the spectral fitting of the unresolved emission (Table 2). 
This also confirms that a substantial proportion ($\approx 70$\%) 
of the unresolved emission is indeed due to truly diffuse gas 
rather than faint point-like sources.

\begin{table}
\caption{Luminosity of emission from 
   discrete and unresolved sources in the nuclear region (0.3--8.0 keV band).}
\label{mathmode} 
\centering 
\begin{tabular}{@{}lrr} 
\hline
\hline \\ 
  luminosity ($\times 10^{38}$~erg~s$^{-1}$) & inside 7\arcsec & inside 16\arcsec \\ [5pt] 
\hline
\hline \\
discrete sources  & $7.7$ & $12.3$\\[5pt]
unresolved sources  & $8.0$ & $11.5$\\[5pt]
{\hspace*{0.3cm}}optically-thin thermal component   & $5.5$ & $7.8$\\[5pt] 
{\hspace*{0.3cm}}power-law  component    & $2.5$ & $3.7$\\[5pt] 
\hline \\ 
total      & $15.7$ & $23.8$\\[5pt]
\hline \\ 
\end{tabular}   
\end{table}

\section{Spectral properties of three bright sources}  

The true-colour X-ray image of the nuclear region (Figure~2, bottom panel)
  and the soft- and hard-band images in Figure~4   
  clearly show the variations of spectral properties 
  among the discrete sources 
  and the different spatial distributions 
  for the soft and hard emission.   
Although a more detailed analysis  
of the discrete sources will be presented  
  in a future paper (Soria et al., in preparation), 
  here we briefly discuss the spectral properties 
  of the two brightest nuclear sources 
  (No.~43 and 44 in Table A.1). 
For comparison, we also present the {\it Chandra} spectrum 
of the brightest source in the galaxy, 
  the {\em ROSAT} HRI source H30 
  (an off-centre source located outside the S3 field of view).   
  
\subsection{Two bright nuclear sources}

Sources No.~43 and 44 are the two brightest nuclear sources 
in the 0.3--8.0~keV band (Table A.1).   
The IR photometric nucleus (source No.~43) is embedded in strong diffuse emission 
from optically thin plasma, which reduces the precision with which 
we can determine its spectral parameters.
We extracted the source and the background spectra 
  using the routine {\it psextract} in {\small{CIAO}}  
  and fitted the background-subtracted spectrum (Figure 7) 
  using standard models in {\small{XSPEC}}.     
The spectrum is well fitted ($\chi^2_\nu = 0.8$, 26~dof) 
by an absorbed power-law model, 
  with a total column density  
  $n_{\rm H} = (1.0^{+1.4}_{-0.6}) \times 10^{21}$ cm$^{-2}$ and a  
  power-law photon index $\Gamma = (1.15^{+0.18}_{-0.22})$ (Table~3). 
The implied emitted luminosity for a distance of 3.7 Mpc 
  is $L_{\rm x} = 2.6 \times 10^{38}$~erg~s$^{-1}$ in the 0.3--8.0~keV band,  
  which is slightly higher than the Eddington luminosity 
  of a 1.5-M$_\odot$ accreting compact object.  
   
Source No.~44 has the highest count rate among all discrete nuclear sources. 
It is located about 7\arcsec~South of the IR photometric nucleus, 
  and is near the southern end of the star-forming ring.   
The source and background spectra were again 
extracted with {\it psextract}.  
A simple absorbed power-law model provides a good fit 
($\chi^2_\nu = 1.02$, 53~dof), with total column density
$n_{\rm H} = (1.9^{+0.4}_{-0.4}) \times 10^{21}$ cm$^{-2}$ and 
power-law photon index $\Gamma = (2.7^{+0.2}_{-0.2})$ (Figure 8 and Table 4).
The deduced emitted luminosity is 
  $L_{\rm x} \simeq 4.6 \times 10^{38}$ in the 0.3--8.0 keV band.

In fact, this choice of spectral model 
may over-estimate the true column density and luminosity, 
if the spectrum turns over at low energies.
Therefore, we also fitted the spectrum with the bmc model 
in {\small{XSPEC}}, which consists of a 
black-body component and a hard, power-law-like tail produced 
by Comptonisation of the soft photons by high energy electrons  
(eg, Shrader \& Titarchuk 1999). 
We obtain an equally good fit ($\chi^2_\nu = 1.02$, 52~dof) with 
the following parameters: total column density 
$n_{\rm H} = (0.9 \pm 0.1) \times 10^{21}$ cm$^{-2}$, 
temperature of the seed photons $kT = (0.16 \pm 0.08)$~keV, 
photon index of the power-law tail $\Gamma = (2.6^{+0.1}_{-0.1})$. 
For this model, the luminosity emitted in the 0.3--8.0 keV band 
is only $L_{\rm x} \simeq 2.4 \times 10^{38}$.

\begin{figure}
\psfig{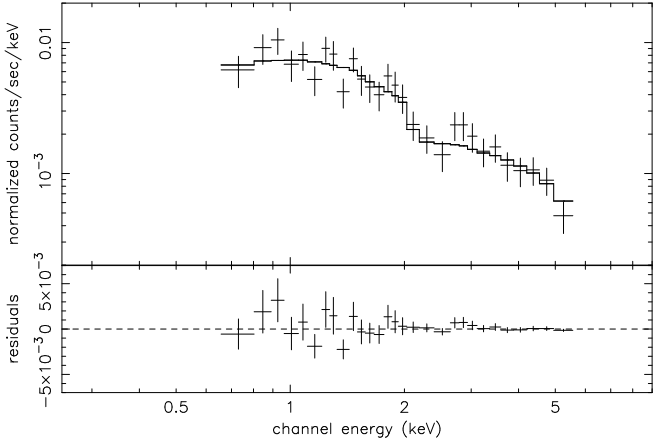}         
\caption{
   Spectrum of the IR photometric nucleus (source No.~43 in Figure 4 and Table A.1).  
   The model spectrum (solid line) is an absorbed power law, 
      and the best-fit parameters are listed in Table 3. 
The background-subtracted spectrum has been grouped 
to achieve a S/N ratio $\geq 4$ in each bin.
}
\label{}
\end{figure}

\begin{figure}
\psfig{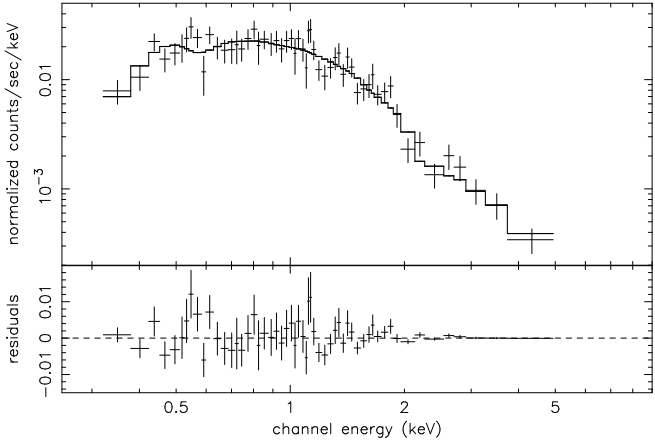}         
\caption{
   Spectrum of source No.~44 in the nuclear region.  
   The model spectrum (solid line) 
      is an absorbed power law, 
      and the best-fit parameters are listed in Table 4. 
The background-subtracted spectrum has been grouped 
to achieve a S/N ratio $\geq 4$.
}
\label{}
\end{figure}

\begin{figure}
\psfig{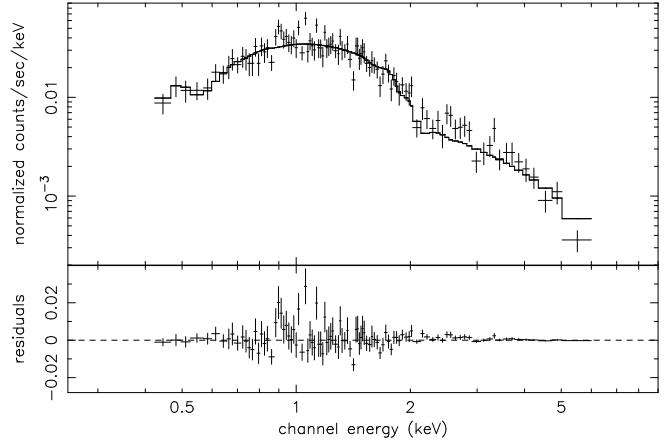}         
\caption{Spectrum of the {\em ROSAT} HRI source H30.  
   The model spectrum (solid line) 
      is an absorbed power law, 
      and the best-fit parameters are listed in Table 5.  
The background-subtracted spectrum has been grouped 
to achieve a S/N ratio $\geq 4$.
}\label{}
\end{figure}

\begin{figure}
\psfig{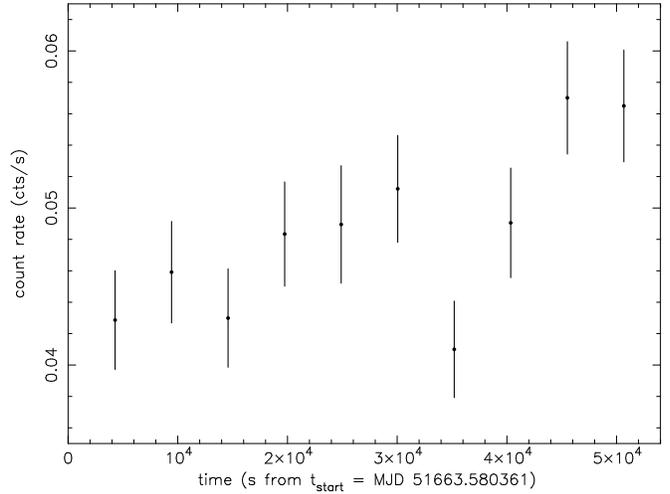}         
\caption{ 
   The 0.3--8.0~keV band light-curve of the {\em ROSAT} HRI source H30 
   shows variability significant at the 90\% level. 
   The data are binned in 10 time intervals of approximately 5~ks each. 
}\label{}
\end{figure}

\subsection{ROSAT source H30}

The source H30 is the brightest point source 
  inside the M83 D$_{25}$ ellipse;  
  it is located at 
{R.A.~(2000) $=$ 13$^h$\,37$^m$\,19$^s$.8}, 
{Dec.~(2000) $=$ $-$29$^{\circ}$\,53\arcmin\,48\farcs9}, 
which is about 4\farcm5 South-East of the nucleus.  
The source is in the S2 chip and therefore is not listed in Table A.1.  

A model with a simple absorbed power law provides an acceptable fit 
($\chi^2_{\nu} = 1.32$, 98~dof), with total absorption column density  
  $n_{\rm H} = (2.5^{+0.4}_{-0.3}) \times 10^{21}$ cm$^{-2}$ and 
  power-law photon index $\Gamma = 2.45^{+0.13}_{-0.12}$  
  (Figure~9 and Table~5).
The implied emitted luminosity 
  is $L_{\rm x} \simeq 11 \times 10^{38}$~erg~s$^{-1}$  
  in the 0.3--8.0 keV band.

As before, lower luminosities are obtained when 
the spectrum is fitted with a Comptonisation model including 
a soft blackbody component and a hard tail. 
Using bmc in {\small{XSPEC}}, we obtain an equally good fit 
($\chi^2_{\nu} = 1.32$, 98~dof) with total 
$n_{\rm H} = (4 \times 10^{20})$ cm$^{-2}$ (fixed at the Galactic value 
with no intrinsic absorption), 
blackbody temperature $kT = (0.26^{+0.01}_{-0.01})$~keV, and 
hard tail photon index $\Gamma = 2.45^{+0.06}_{-0.08}$  
(Table~5). For this choice of model, 
the implied emitted luminosity 
  is $L_{\rm x} \simeq 5.6 \times 10^{38}$~erg~s$^{-1}$  
  in the 0.3--8.0 keV band.

As a test, we then fitted the spectrum with a disk blackbody model, 
which also provides an acceptable fit ($\chi^2_{\nu} = 1.33$, 99~dof). 
The best-fit disk-blackbody colour temperature 
$kT_{\rm in} = (0.92^{+0.05}_{-0.04})$~keV, 
the emitted luminosity $L_{\rm x} \simeq 5.4 \times 10^{38}$~erg~s$^{-1}$,  
and the other parameters are listed in Table~5. 

Finally, we fitted the spectrum with 
a thermal bremsstrahlung model,  
which formally provides the best fit ($\chi^2_{\nu} = 1.23$, 99~dof). 
The fit parameters (Table~5) are in agreement, at 2-$\sigma$ level, 
  with those obtained by Immler et al.\ (1999). 
With this model, the emitted luminosity in the 0.3--8.0~keV band  
  is $L_{\rm x} \simeq 6.7 \times 10^{38}$ erg s$^{-1}$,    
  corresponding to $L_{\rm x} \simeq 5.5 \times 10^{38}$ erg s$^{-1}$
in the 0.1--2.4~keV band and $L_{\rm x} \simeq 4.2 \times 10^{38}$ erg s$^{-1}$
in the 0.5--3.0~keV band.
This is roughly consistent with the values of 
$6.7 \times 10^{38}$ erg s$^{-1}$ obtained from the {\em ROSAT} 
HRI data at 0.1--2.4~keV, 
  $5.7 \times 10^{38}$ erg s$^{-1}$ obtained from 
the {\em Einstein} IPC data at 0.5--3.0~keV, and 
  $4.2 \times 10^{38}$ erg s$^{-1}$ obtained from 
the {\em Einstein} HRI data also at 0.5--3.0~keV,
for a distance of 3.7~Mpc   
  (Immler et al.\ 1999;  Trinchieri et al.\ 1985). 
We therefore conclude that the source  
  does not show strong evidence of long-term X-ray variability. 

The source appears to show an increase in brightness 
during the 50-ks {\em Chandra} observation (Figure 10).    
Fitting a constant to the count rate, binned into ten 5-ks time intervals, 
  yields $\chi_\nu^2 = 2.7$ (9~dof),   
  while a linear fit with $dS/dt = 2.4 \times 10^{-7}$ cts s$^{-2}$  
  gives $\chi_\nu^2 = 1.6$ (8~dof).  
The $\Delta \chi^2$ indicates 
  that the variability is significant at the 90\% level. 
Short-term variabilities were also detected 
  during the {\em ROSAT} observations (Immler et al.\ 1999).    

\begin{table}
\caption{
   {\small XSPEC} best-fit parameters 
   for the X-ray source at the IR photometric nucleus (No.~43 in Table A.1). 
The foreground (Galactic) line-of-sight absorption column density has been fixed at 
    $n_{\rm H, Gal} = 4 \times 10^{20}$~cm$^{-2}$.}
\label{mathmode} 
\centering 
\begin{tabular}{@{}lr} 
\hline
\hline \\ 
\multicolumn{2}{c}{model: tbabs$_{\rm Gal}$ $\times$ tbabs $\times$ power-law  }\\[5pt]  
\hline
\hline \\ 
$n_{\rm H}~(\times 10^{20}$~cm$^{-2})$  & $5.6^{+14.2}_{-5.6}$   \\[5pt] 
$\Gamma$       & $1.15^{+0.18}_{-0.22}$ \\[5pt]
$K_{\rm pl}~(\times 10^{-5})$    & $1.5^{+0.5}_{-0.3}$ \\ [5pt]   
\hline \\ 
$\chi_\nu^2$~(dof) &  0.80~(26)  \\ [5pt] 
$L_{\rm{0.3-8}}~(\times 10^{38}$~erg~s$^{-1})$ & 2.6  \\[5pt]
\hline \\
\end{tabular}   
\end{table}

\begin{table}
\caption{
   {\small XSPEC} best-fit parameters 
   for the nuclear source No.~44.
   The foreground column density has been fixed at 
    $n_{\rm H, Gal} = 4 \times 10^{20}$~cm$^{-2}$. }
\label{mathmode} 
\centering 
\begin{tabular}{@{}lr} 
\hline
\hline \\  
\multicolumn{2}{c}{model: tbabs$_{\rm Gal}$ $\times$ tbabs $\times$ power-law}  
	 \\[5pt]
\hline
\hline \\
$n_{\rm H}~(\times 10^{20}$ cm$^{-2})$     & $18.9^{+4.2}_{-4.0}$  \\[5pt]
$\Gamma$        & $2.74^{+0.12}_{-0.20}$ \\[5pt]
$K_{\rm pl}~(\times 10^{-5})$       & $5.9^{+1.0}_{-0.9}$  \\ [5pt]   
\hline \\  
 $\chi_\nu^2$~(dof)  &  1.02~(53)   \\ [5pt] 
 $L_{\rm{0.3-8}}~(\times 10^{38}$~erg~s$^{-1})$ & 4.6  \\ [5pt]
\hline
\hline \\  
\multicolumn{2}{c}{model: tbabs$_{\rm Gal}$ $\times$ tbabs $\times$ bmc}  
	 \\[5pt]
\hline
\hline \\
$n_{\rm H}~(\times 10^{20}$ cm$^{-2})$     & $4.9^{+0.9}_{-0.9}$  \\[5pt]
$T_{\rm bb}$~(keV)  & $0.16^{+0.08}_{-0.08}$\\[5pt]             
$\Gamma$        & $2.59^{+0.14}_{-0.14}$ \\[5pt]
$K_{\rm bmc}~(\times 10^{-6})$       & $1.1^{+0.6}_{-0.6}$  \\ [5pt]   
\hline \\  
 $\chi_\nu^2$~(dof)  &  1.02~(52)   \\ [5pt] 
 $L_{\rm{0.3-8}}~(\times 10^{38}$~erg~s$^{-1})$ & 2.4  \\ [5pt]
\hline
\end{tabular}   
\end{table}

\begin{table}
\caption{
   {\small XSPEC} best-fit parameters for the {\em ROSAT} HRI source H30. 
    The foreground column density has been fixed at 
    $n_{\rm H, Gal} = 4 \times 10^{20}$~cm$^{-2}$.}
\label{mathmode} 
\centering 
\begin{tabular}{@{}lr} 
\hline
\hline \\ 
\multicolumn{2}{c}{model: tbabs$_{\rm Gal}$ $\times$ tbabs $\times$ power-law}\\[5pt]  	
\hline 
\hline \\ 
$n_{\rm H}~(\times 10^{21}$~cm$^{-2})$ & $2.1^{+0.4}_{-0.3}$ \\[5pt]
$\Gamma$        & $2.45^{+0.13}_{-0.12}$ \\[5pt]
$K_{\rm pl}~(\times 10^{-4})$   & $1.38^{+0.16}_{-0.14}$ \\ [5pt]   
\hline \\ 
$\chi_\nu^2$~(dof)  &  1.32~(98)  \\[5pt] 
$L_{\rm{0.3-8}}~(\times 10^{38}$~erg~s$^{-1})$ & 11  \\[5pt]
\hline
\hline \\ 
\multicolumn{2}{c}{model: tbabs$_{\rm Gal}$ $\times$ tbabs $\times$ diskbb}\\[5pt]  	
\hline 
\hline \\ 
$n_{\rm H}~(\times 10^{21}$~cm$^{-2})$ (fixed) & $0.0$ \\[5pt]
$T_{\rm in}$~(keV) & $0.92^{+0.05}_{-0.04}$ \\ [5pt]             
$K_{\rm dbb}~(\times 10^{-2})$ & $2.25^{+0.47}_{-0.38}$ \\[5pt] 
\hline \\ 
$\chi_\nu^2$~(dof)  &  1.33~(99)  \\[5pt] 
$L_{\rm{0.3-8}}~(\times 10^{38}$~erg~s$^{-1})$ & 5.4  \\[5pt]
\hline
\hline \\
\multicolumn{2}{c}{model: tbabs$_{\rm Gal}$ $\times$ tbabs $\times$ bmc}\\[5pt]  	
\hline 
\hline \\ 
$n_{\rm H}~(\times 10^{21}$~cm$^{-2})$ (fixed) & $0.0$ \\[5pt]
$T_{\rm bb}$~(keV) & $0.26^{+0.01}_{-0.01}$ \\ [5pt]             
$\Gamma$         & $2.45^{+0.06}_{-0.08}$ \\[5pt]
$K_{\rm bmc}~(\times 10^{-6})$ & $2.4^{+0.2}_{-0.2}$ \\[5pt]  
\hline \\ 
$\chi_\nu^2$~(dof)  &  1.32~(98)  \\[5pt] 
$L_{\rm{0.3-8}}~(\times 10^{38}$~erg~s$^{-1})$ & 5.6  \\[5pt]
\hline
\hline \\
\multicolumn{2}{c}{model: tbabs$_{\rm Gal}$ $\times$ tbabs $\times$ bremsstrahlung }\\[5pt]    
\hline 
\hline \\ 
$n_{\rm H}~(\times 10^{21}$~cm$^{-2})$ & $0.82^{+0.23}_{-0.22}$ \\ [5pt]
$T_{\rm br}$~(keV) & $2.56^{+0.32}_{-0.28}$ \\[5pt]
$K_{\rm br}~(\times 10^{-4})$  & $1.29^{+0.13}_{-0.11}$ \\[5pt]
\hline \\ 
$\chi_\nu^2$~(dof) &  1.23~(98) \\[5pt]
$L_{\rm{0.3-8}}~(\times 10^{38}$~erg~s$^{-1})$ & 6.7  \\[5pt]
\hline
\end{tabular}   
\end{table}  

\section{Discussion}  
 
\subsection{Luminosity distribution of the sources}    

Recent observations have shown that 
  the luminosity distributions of X-ray sources 
  in nearby galaxies can often be approximated by a power-law   
  (eg, disk sources in M101, Pence et al.\ 2001, 
  and M81, Tennant et al.\ 2001)
  or by a broken power-law profile
  (eg, bulge sources in M31, Shirey et al.\ 2001, 
and in M81, Tennant et al.\ 2001).    
The log~N($>$S) -- log~S curves for the sources located 
in the bulges of spiral galaxies  
  tend to have steeper slopes at the high-luminosity end. 
For example, the broken power-law distribution for bulge sources 
in M31 has slopes of $-0.5$ and $-1.8$ at the low- and high-luminosity 
end, respectively (Shirey et al.\ 2001).  
The log~N($>$S) -- log~S curves for the sources in galactic disks   
  are generally single, flatter power laws 
  (eg, with a slope of $-0.5$ for the disk sources in M81, 
Tennant et al.\ 2001, and $-0.8$ for the disk sources in M101, Pence et al.\ 2001). 
Elliptical galaxies also have broken power-law log~N($>$S) -- log~S curves, 
  with generally steep slopes at their bright ends 
  (eg, with a slope of $-1.8$ for NGC~4697, 
  Sarazin, Irwin \& Bregman 2000), 
  similar to those inferred for the bulge sources in spiral galaxies.  
Starburst galaxies, instead,  
  tend to have flat power-law log~N($>$S) -- log~S curves 
  (eg, with a slope of $-0.45$ for NGC~4038, Fabbiano, Zezas \& Murray 2001) 
  similar to the distributions for the disk sources in spiral galaxies.  
     
A flatter power-law luminosity distribution    
  implies a larger proportion of bright sources in a population. 
If the brightest X-ray sources in a galaxy 
  are young, short-lived high-mass X-ray binaries, 
  born in a recent starburst episode,    
  then the slope of the bright end of the log~N($>$S) -- log~S curve 
  indicates the star-formation activity of the galaxy (see Prestwich 2001).     

Various mechanisms can produce a broken power-law profile 
in the log~N($>$S) -- log~S curve.  
The break may be caused by a pile-up of systems 
at a particular luminosity;  
in particular, it may be due to a population of neutron-star X-ray binaries  
(Sarazin, Irwin \& Bregman 2000) 
with a mass-transfer rate at or just above the Eddington limit, i.e.,   
with bolometric luminosities $\sim 2 \times 10^{38}$~erg~s$^{-1}$.    
Aging of a population of X-ray binaries born during a starburst episode  
  can also produce a luminosity break in the log~N($>$S) -- log~S curve 
  (Wu 2001; Wu et al.\ 2001). In this model,   
the initial distribution has a power-law profile; 
  as the bright, short-lived systems die, 
  a break is created, moving gradually to lower luminosities with time.
Pile-up of neutron-star X-ray binaries 
  is a likely cause of the luminosity break 
  in the log~N($>$S) -- log~S curves of elliptical galaxies, 
  where active star formation is absent.   
Population aging is a more likely mechanism for spiral galaxies,  
  in particular those that have had tidal interactions 
with their satellites in the recent past 
  (Wu 2001; Wu et al.\ 2001).

We have shown in Section 3 that the sources in the nuclear region of M83 
and those in the disk appear to have different luminosity distributions. 
One of the obvious features in the log~N($>$S) -- log~S curve 
  of the disk sources is a kink,  
  located roughly at the Eddington luminosity of accreting neutron stars   
  (if a distance of 3.7~Mpc is assumed). The slopes are 
approximately $-0.6$ and $-1.3$ at the low- and high-luminosity end, 
respectively.
When we consider only the sources inside the 60\arcsec~central region 
(which includes the nucleus and the stellar bar, but not the spiral arms), 
we obtain a simple power-law distribution with a slope of $-0.8$ (Figure 3).
This implies that 
  the population of sources in the inner regions (nucleus and bar)
  has a larger relative fraction of bright sources than 
the disk population.

The situation is different for example 
  in the spiral galaxy M81, 
  where most bright sources are found in the galactic disk 
  instead of the nuclear region (Tennant et al.\ 2001).  
If the flatness of the slope in the log~N($>$S) -- log~S curve 
  is an indicator of recent star formation, 
  the difference in the spatial distribution of the brightest sources 
  in M83 and M81 is simply a consequence of the fact that 
  M83 has a starburst nucleus  
  while star formation in galaxies such as M81 is presently more efficient 
  in the galactic disk. With this interpretation, the current star formation rate 
in the disk of M83 would be intermediate between the rate in the disk 
of M81 (slope of $-0.5$ at the high-luminosity end) 
and in the bulge of M31 (slope of $-1.8$). 
 
\subsection{Nuclear region}    

The distribution of bright, young star clusters suggests that 
   the most vigorous star formation  
   is concentrated in a semi-circular annulus 
   $\simeq 7\arcsec$~($\approx 130$ pc) 
   South-West of the IR photometric nucleus.   
From the colour distribution of the star clusters, 
   it is inferred that 
   star formation first started at the southern end of the ringlet  
   about $10$--$30$~Myr~ago  
   (Thatte et al.\ 2000; Harris et al.\ 2001),  
   and has since propagated towards the currently more active northern end, 
   where the age of most young stellar clusters is $\simlt 5$~Myr.  
The youngest clusters are found along the outer edge of the ringlet,
   indicating that the star-formation front is presently propagating outwards  
   (Harris et al.\ 2001). 

The starburst nucleus of M83 was unresolved 
   in all X-ray observations before {\em Chandra};    
   therefore, no comparisons between observations in the X-rays 
and in other wavelengths had been possible.       
Overplotting the {\em Chandra} brightness contours 
in the 0.3--8.0~keV band   
   on the {\em HST}/WFPC2 multi-colour images (Figure~11)  
   helps to shed light on the relative spatial distribution 
of the X-ray and optical emission.    
For instance, although there is a general correlation, the 
optical emission is more strongly 
concentrated around the IR photometric nucleus and along 
the star-forming ring, while the X-ray emission  
is more uniformly distributed. 
There is also extended X-ray emission 
   South-West of the star-forming arc, 
   along the direction of the main galactic bar, 
   and towards the North-West, across the dust lane, 
   in a direction perpendicular to the bar.    
Discrete X-ray sources and unresolved emission are also found   
to the East of the IR photometric nucleus, 
   not associated with any bright optical regions 
   with currently active star formation. 
This may be due to the fact that the 
   optical emission traces the bright young stellar clusters 
and OB associations, 
   while the X-ray emission is in general associated  
   with remnants of stellar evolution 
   such as accreting compact stars and supernovae.  

The HST image shows that the IR photometric nucleus 
is strongly extincted in the UV.     
The X-ray spectrum of the corresponding {\it Chandra} 
source (No.~43 in Table A.1) does not allow a precise determination 
of the absorption column density, but it constrains 
it to be $< 2.4 \times 10^{21}$~cm$^{-2}$ (Table 3). 
From the relation between the absorption column density $n_{\rm H}$ 
  and the visual extinction $A_V$ (Predehl \& Schmitt 1995),   
  this implies that the source has a visual extinction $A_V < 1.3$ mag.   
An extinction $A_V = 0.9$ mag has been deduced from IR observations
  (Thatte et al.\ 2000). 

Unresolved soft emission extends for $\simgt 15$\arcsec~ 
   ($\simgt 270$~pc) to the South-West, West and North-West, 
   outside the star-forming arc; 
   the unresolved hard X-ray emission appears instead 
   to be confined in the region  
   between the IR photometric nucleus and the star-forming arc 
(Figure 4, bottom right panel),  
   and may perhaps extend to the North-West for $\simlt 10$\arcsec~ 
   ($\approx 180$ pc).   
The X-ray spectrum of the unresolved component   
   shows strong emission lines, 
   typical of emission from optically-thin thermal plasma 
at $kT \approx 0.6$ keV.  Above-solar abundances of Ne, Mg, Si and S 
are required to fit the spectrum, while Fe appears to be underabundant. 
This suggests that the interstellar 
medium in the starburst nuclear region has been enriched by the ejecta of type-II 
supernova explosions. A high C abundance and a high C/O abundance ratio  
can be the effect of radiatively-driven winds from metal-rich massive stars 
($M \simgt 40$\,M$_{\odot}$) in their Wolf-Rayet stage (Gustafsson et al 1998; 
Portinari, Chiosi \& Bressan 1998). Both effects are likely to be present 
in the nuclear region.  

Four discrete sources are resolved along the star-forming arc. 
The brightest of them (No.~44 in Table A.1) is located near the southern end of the arc, 
   where star formation started $\approx 10$--$30$~Myr~ago 
   (Harris et al.\ 2001).
The other three sources (No.~35, 37 and 40) are towards 
the northern end of the arc, 
    where star formation started more recently ($\simlt 5$ Myr ago); 
    source No.~35 is approximately coincident with the location of SN~1968L, 
   a type-II supernova (Wood \& Andrews 1974).

                             
\begin{figure*}
\vbox{
\begin{tabular}{lr}
\psfig{figure=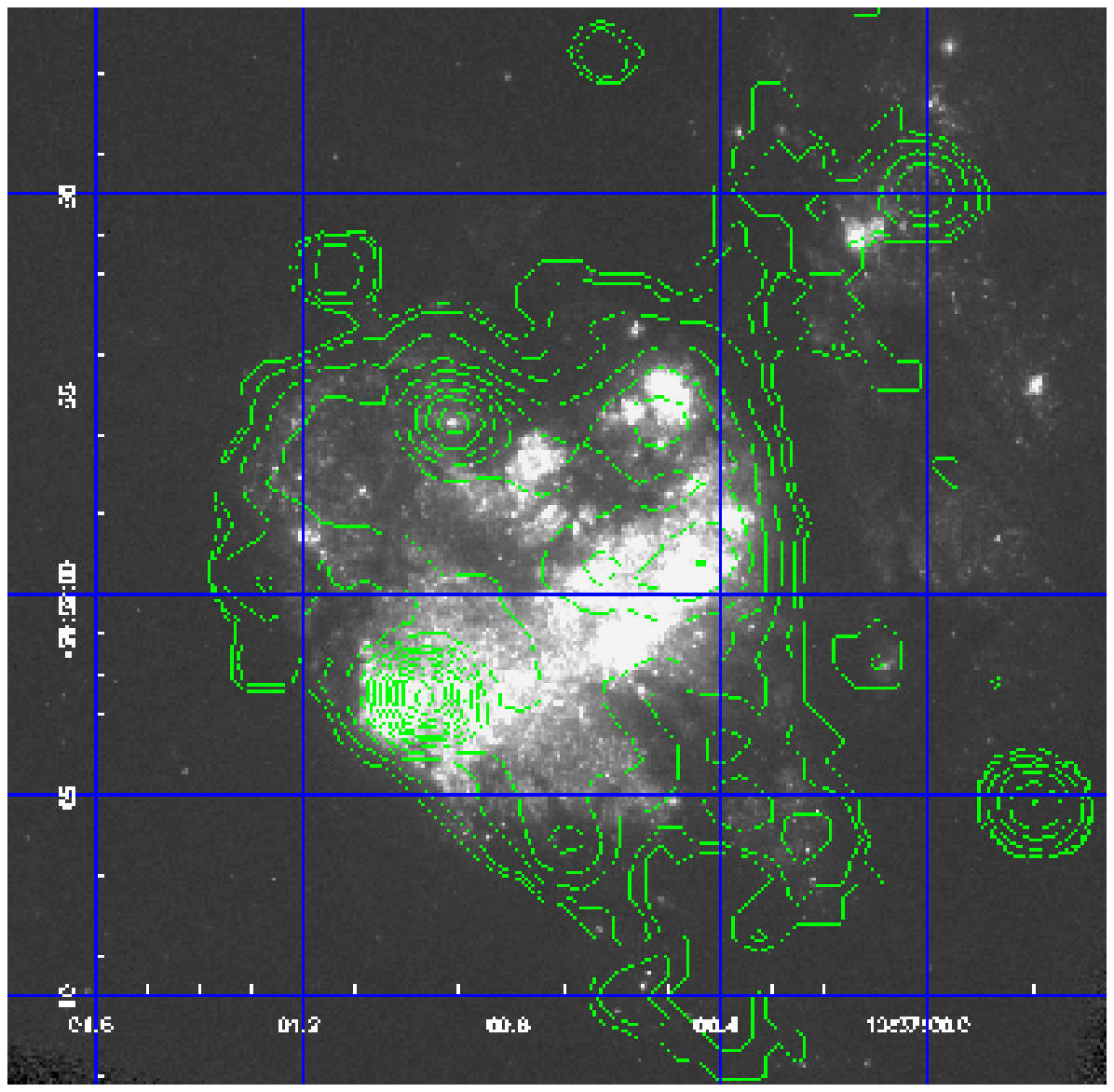,width=8.5cm} &
\psfig{figure=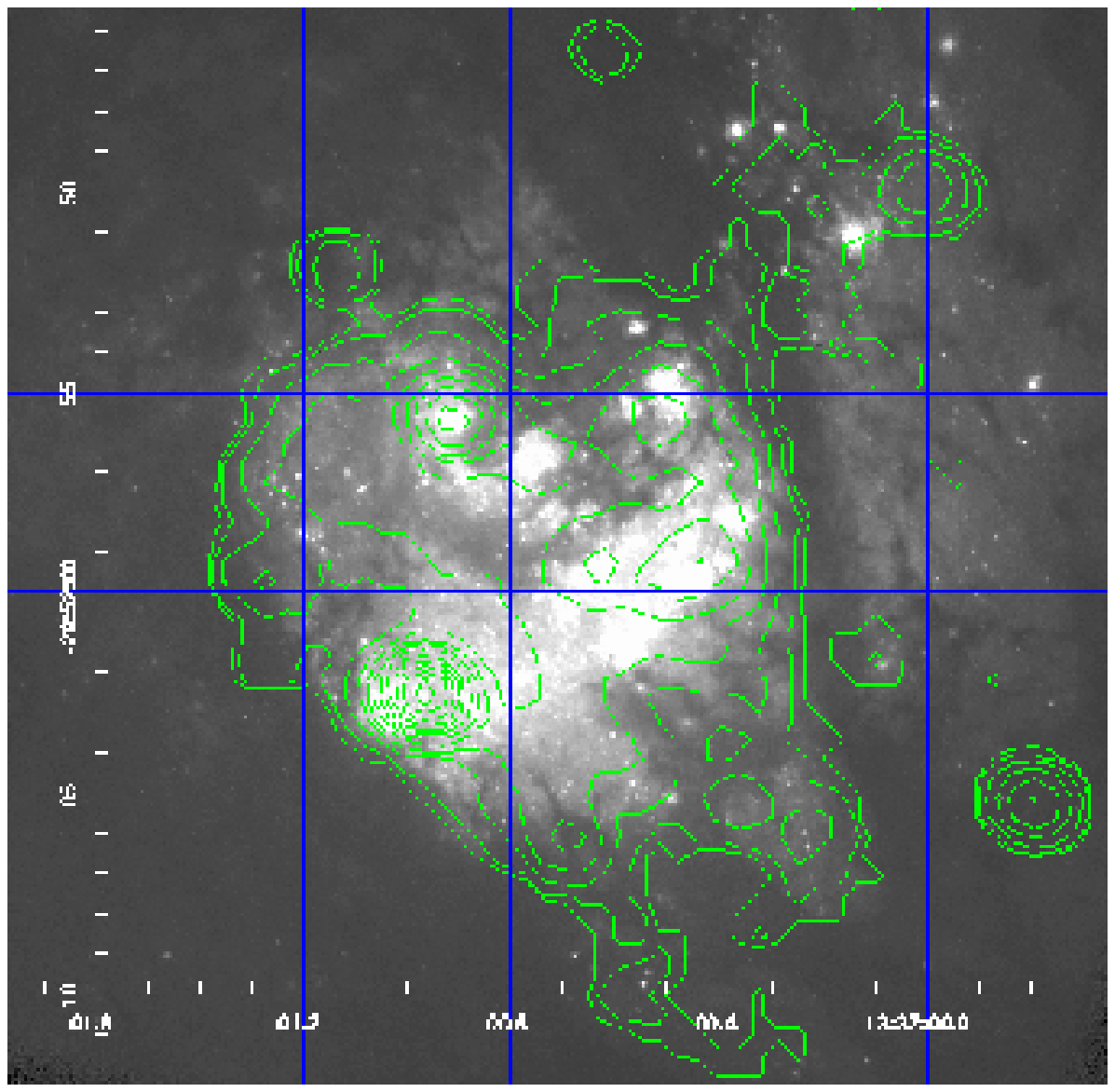,width=8.5cm} \\
\psfig{figure=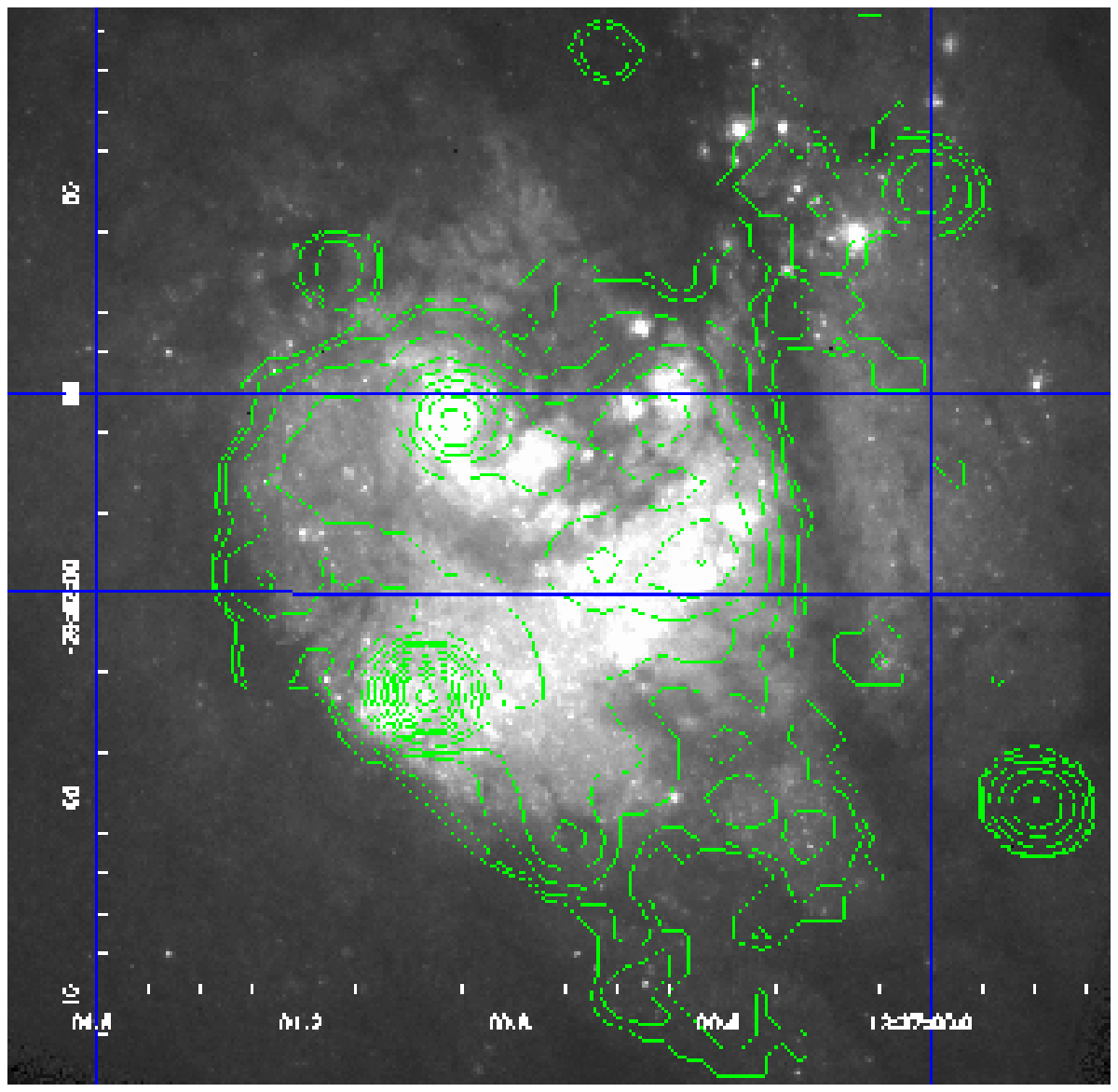,width=8.5cm} &
\psfig{figure=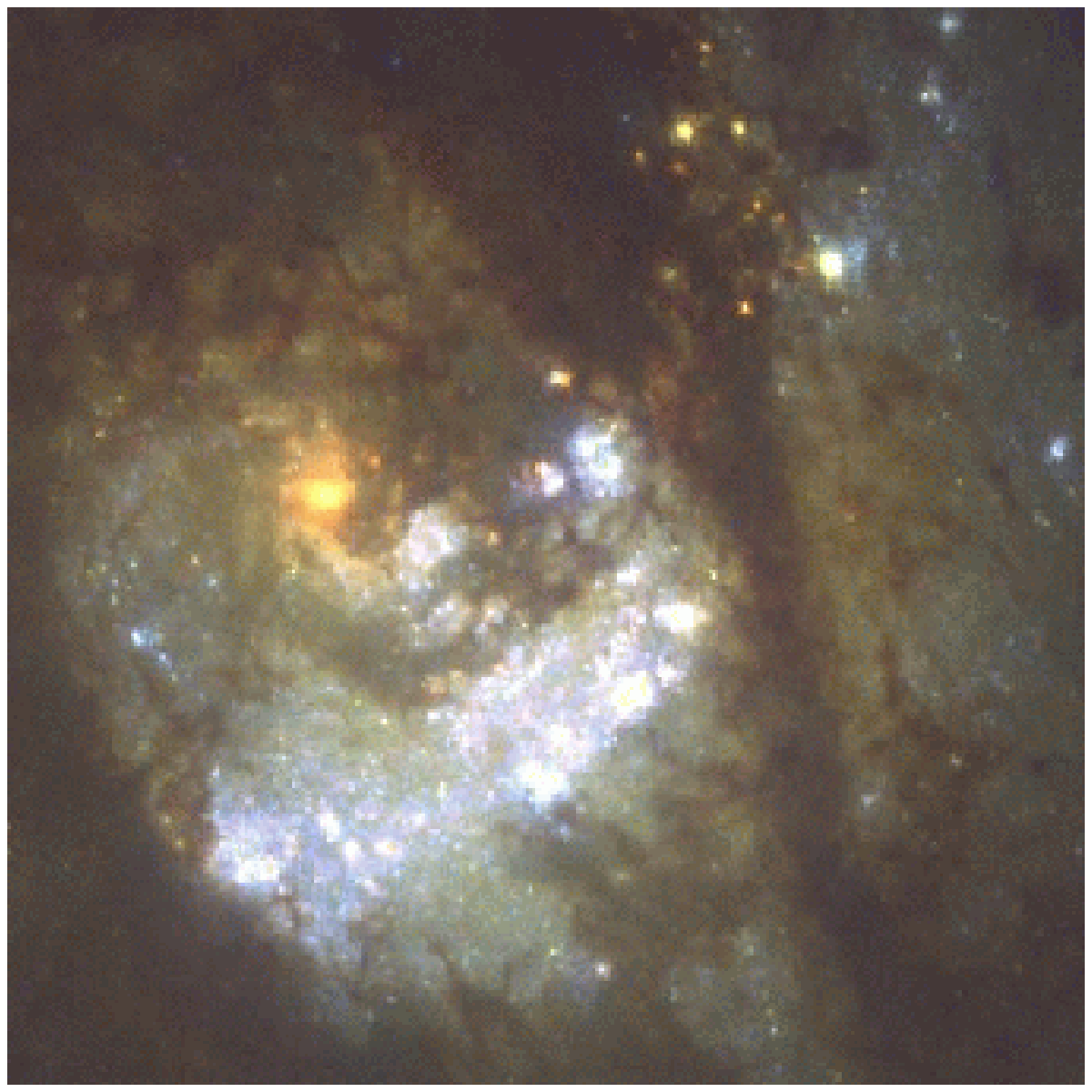,width=8.35cm} \\
\end{tabular}
}
\caption{The contours of the 0.3--8.0 keV emission in the nuclear region 
     (smoothed with a $1\farcs5 \times 1\farcs5$ boxcar kernel)  
     are overplotted on the {\em HST}/WFPC2 images 
     in the F300W ($\approx U$; top left panel), F547M ($\approx V$; top right) 
     and F814W ($\approx I$; bottom left) broad-band filters. 
  A three-colour {\em HST}/WFPC2 image is also shown for comparison (bottom right,   
     from Harris et al.\ 2001).
  The HST images, taken on 2000 April--May,
     were obtained from the STScI public archive. North is up, East is left. 
}
\label{fig:image}
\end{figure*}

\subsection{Black-hole candidates}  

Active galactic nuclei (AGN) are often found to have power-law X-ray spectra. 
If a thermal blackbody component is present, 
  it peaks at energies $\simlt 0.1$ keV 
  and does not dominate in the 0.3--8.0~keV band.   
Black-hole X-ray binaries in hard spectral states 
  also have single power-law spectra in the 0.3--8.0~keV band. 
  When they are in the soft spectral state, instead, 
  their X-ray spectra consist of a strong thermal blackbody component, 
  with a temperature of about 1~keV,   
  and a power-law tail, usually steeper than the power law in the hard state.

The IR photometric nucleus has a power-law X-ray spectrum  
  consistent with either those of accreting supermassive black holes in AGN, 
  or those of stellar-mass black-hole candidates in X-ray binaries, 
in the hard state.     
The mass associated to the source is estimated, from stellar kinematics, 
  to be $\approx 10^7$~M$_{\odot}$ (Thatte et al.\ 2000). 
If the source contains a supermassive black hole, 
analogous to those found in the nuclei of common Seyferts and quasars, 
and if the X-ray emission is powered by accretion, 
its luminosity 
  ($L_{\rm x} \simeq 3.2 \times 10^{38}$~erg~s$^{-1}$ 
in the 0.3--8.0~keV band) is well below the Eddington limit; 
therefore, M83 is not in an AGN phase in the present epoch.

In the most accepted scenario, short-lived massive stars
   are the progenitors of stellar-mass black-hole candidates. Thus, 
   some of the discrete X-ray sources in the nuclear region 
   (excluding possible supernova remnants and the IR nucleus itself)  
   may be high-mass black-hole X-ray binaries. 
Source No.~44 is a likely black-hole candidate, end product of 
the most recent starburst episode in the region (age $< 30$ Myr): 
its unabsorbed luminosity $\simgt 2.4 \times 10^{38}$~erg~s$^{-1}$ 
  in the 0.3--8.0~keV band places it above 
  the luminosity of a neutron star accreting material 
at the Eddington rate.

Although the two brightest nuclear sources, No.~43 and 44, 
  have comparable luminosity,   
  and they are probably both powered by accretion onto a black hole, 
  their spectral properties are different.  
The hard power-law photon index ($\Gamma < 1.5$) in the spectrum 
of the photometric nucleus
  and the kinematic properties of the stars around it 
  suggest that it is a supermassive black hole. 
The other source has a much softer power-law photon index ($\Gamma > 2.5$) 
consistent with the spectrum of a stellar-mass black-hole candidate 
in the soft state. The present data are not sufficient to determine  
whether or not a disk-blackbody component is also present.

We can compare these two nuclear black-hole candidates 
with the brightest X-ray source in the M83 field, 
the {\em ROSAT} HRI source H30, which is located 
$\approx 4\farcm5$ ($\sim 5$ kpc) away from the nucleus. 
Immler et al.\ (1999) suggested 
  that H30 is also a strong black-hole candidate,  
  because its luminosity is well above the Eddington limit  
  for a 1.5-M$_\odot$ compact star. 
(The source is still super-Eddington 
  after accounting for the smaller distance to M83 assumed in this work.)   
Its spectrum can be fitted equally well with various models (Table 5). 
The most likely physical interpretation is that it 
is dominated by a steep ($\Gamma \approx 2.5$) power-law component, 
produced by Comptonisation of a soft blackbody component. 
The current data do not allow a simultaneous determination 
of the absorbing column density and of the temperature of the 
blackbody component. If the blackbody temperature 
$T_{\rm bb} \ll 0.2$ keV, the total column density 
$n_{\rm H} \approx 2.5 \times 10^{21}$~cm$^{-2}$. 
In the other extreme case, for $n_{\rm H} = 4 \times 10^{20}$~cm$^{-2}$ 
(no intrinsic absorption), $T_{\rm bb} \approx 0.25$ keV. 
An accurate determination of the optical extinction of a possible optical 
counterpart would constrain the temperature of the blackbody component in the 
X-ray spectrum, and hence the mass of the accreting object. This is left 
to further work.

The spectrum of H30 can also be well fitted 
  with an optically-thin thermal plasma model 
(bremsstrahlung emission from completely ionised plasma at 
$kT \approx 2.5$ keV), consistent with a luminous supernova remnant.
However, this latter possibility is much more unlikely 
because the source has shown variability 
during the {\em ROSAT} HRI observations 
and over the timespan of the {\em Chandra} observations. 
Moreover, no X-ray emission lines are detected in the spectrum, 
but the best-fit temperature is too low for 
the plasma to be completely collisionally-ionised. This would imply 
a metal abundance $\simlt 0.1$ times the solar value (we have 
estimated it by fitting an absorbed Raymond-Smith model in {\small {XSPEC}}).
Alternatively, the gas could be photo-ionised.
Finally, a simple disk-blackbody spectrum with colour temperature 
of the innner accretion disk 
$T_{\rm in} \approx 0.9$ keV, although physically unlikely, 
cannot be ruled out with the current data.

\section{Summary}

We analysed ACIS S3 data of the {\em Chandra} observation of M83.  
The starburst nuclear region is resolved in X-rays for the first time. 
Eighty-one point sources are detected above 3.5-$\sigma$.  
The sources are highly concentrated in the nuclear region, 
  and 15 of them are found within the inner 16\arcsec~region.  
One strong source coincides with the IR photometric peak,  
  which is believed to be one of the two dynamical nuclei of the galaxy.   
We detect unresolved emission but no point-like sources 
(at a 2.5-$\sigma$ level) 
  at the centre of the outer optical isophote ellipses, 
where another dynamical nucleus is suspected to be located.   

About 50\% of the total emission in the nuclear region  
  is resolved into discrete sources.
The unresolved emission is extended outside  
the star-forming arc, both along the direction of the main 
galactic bar and perpendicular to it.
In spite of the asymmetry, 
  the azithumally-averaged radial distribution 
  of the unresolved emission 
  appears to follow a King profile, 
with a central plateau, corresponding to the region 
inside the outer dust ring seen from IR observations, 
and no central cusp.  
The spectrum of unresolved emission 
  shows strong emission lines, 
  characteristic of an optically-thin thermal plasma 
at a temperature of $\approx 0.60$ keV. We estimate 
that $\approx 70$\% of the unresolved emission 
(35\% of the total) is due to truly diffuse plasma, 
with the rest (15\% of the total) coming from 
faint, unresolved point-like sources and photons in the wings 
of the PSF outside the detection cells of the resolved sources.

A much better fit to the diffuse emission is obtained 
when we assume higher metal abundances for selected 
elements:  C (as expected if the interstellar medium 
in the nuclear region is strongly enriched by winds 
from massive Wolf-Rayet stars), Ne, Mg, Si and S 
(as expected if the gas is enriched by supernova 
ejecta). The emission lines are redshifted, 
implying projected radial velocities $\approx 7000$ km s$^{-1}$.

Separating the discrete sources inside and outside a 60\arcsec~central region  
  reveals that the two populations have different cumulative 
luminosity distributions.    
The log~N($>$S) -- log S curve of the sources outside this radius 
(i.e., the disk population) shows a kink at luminosities 
consistent with the Eddington limit for neutron stars, 
indicating that a substantial fraction of the X-ray binaries in the disk 
  contains a neutron star. 
No such feature is seen for the sources inside the 60\arcsec~radius 
(i.e., those located in the nuclear region and along the bar). 
The slope of the log~N($>$S) -- log S curve at its high-luminosity end 
is flatter for the nuclear population, implying a larger fraction 
of bright sources (possible black-hole candidates). We interpret 
this as evidence that star formation is currently more active 
in the nuclear region than in the disk.

\begin{acknowledgements}
     We thank Reiner Beck, Roy Kilgard, 
Miriam Krauss, Casey Law, Oak-Young Park, Elena Pian, 
Allyn Tennant and Daniel Wang for helpful discussions and suggestions. 
We are particularly grateful to the referee (Stefan Immler) 
for his detailed comments, which helped improve the paper substantially.
This work is partially supported by a University of Sydney Sesqui
R \& D Grant.
KW acknowledges support from Australian Research Council
through an Australian Research Fellowship. 
\end{acknowledgements}


\appendix
\section{}

   \begin{table*}
      \caption[]{
    M83 sources detected at 3.5-$\sigma$ level
    in the 0.3--8.0 keV band, and respective counts in a soft, medium and hard band, 
    whenever the sources are also detected at 3-$\sigma$ level in that band.}
         \label{}
   $$
         \begin{array}[width=textwidth]{rcccrrrrl}
            \hline
            \noalign{\smallskip}
           {\mathrm{No.}} & &
          {\mathrm{R.A.(2000)}}     & {\mathrm{Dec.(2000)}}
                 & {\mathrm{net~counts}}^{\mathrm{a}}
		 & {\mathrm{0.3-1~keV}}
		 & {\mathrm{1-2~keV}}
		 & {\mathrm{2-8~keV}}
                 & {\mathrm{notes}} \\
            \noalign{\smallskip}
            \hline
            \noalign{\smallskip}
  1&  &13~36~40.9 & -29~51~11.0~&     64.2\pm12.2    & & 35.0\pm08.6& & \\
  2&  &13~36~41.4 & -29~50~44.6 &     260.9\pm22.4      &  61.1\pm11.9 & 138.3\pm15.0& 62.3\pm14.0 & \\
  3&  &13~36~43.5 & -29~51~06.8 &     350.4\pm22.9    & 52.5\pm09.7 & 169.5\pm16.2& 129.1\pm14.8 & {\rm H08}   \\
  4&  &13~36~48.0~& -29~53~24.5~&      48.3\pm09.2    & & 22.0\pm06.4&  &   \\
  5&  &13~36~49.1~& -29~52~58.3~&     264.8\pm19.4    & 158.2\pm15.5 & 93.1\pm11.9&  & {\rm H12a} \\

  6&  &13~36~49.2~& -29~53~03.4~&     102.5\pm12.6    & 45.9\pm08.7 &  46.3\pm09.1&  & {\rm H12b} \\
  7&  &13~36~49.8~& -29 52 17.3~&      32.3\pm08.5    & 24.2\pm07.5&  &  &  \\
  8&  &13~36~51.7~& -29~53~34.9~&      96.4\pm13.9    & 35.6\pm09.4&  42.9\pm09.5&  21.7\pm07.0& {\rm H14}  \\
  9&  &13~36~53.2~& -29~53~25.2~&      35.4\pm09.7    & 28.8\pm08.6&  &  & \\
 10&  &13~36~53.3~& -29~52~42.6~&      30.6\pm08.4    & 21.3\pm06.9&  &  & \\

 11&  &13~36~53.9 & -29~48~47.9~&      37.4\pm09.8    & 31.2\pm08.8&  &  & \\
 12&  &13~36~53.9~& -29~51~14.5~&      73.0\pm10.6    & 37.9\pm07.9&  26.2\pm06.8&  & \\
 13&  &13~36~55.0~& -29~52~39.4~&      31.1\pm07.8    & 25.4\pm07.2&  &  & \\
 14&  &13~36~55.2~& -29~54~03.6~&      86.7\pm11.6    & 38.0\pm08.0&   35.2\pm07.8&  & \\
 15&  &13~36~55.5~& -29~55~10.1~&     286.3\pm20.9    & 68.3\pm11.6 &   134.9\pm14.3&  83.2\pm12.2& {\rm H15}  \\

 16&  &13~36~55.6~& -29~53~03.5~&      40.1\pm09.5    &  &  &   &  \\
 17&  &13~36~56.2~& -29~52~55.3~&      29.6\pm07.5    &  &   &  &  \\
 18&  &13~36~56.7~& -29~49~12.2~&     287.8\pm21.2    & 78.7\pm11.3 & 137.8\pm14.7  &  71.3\pm12.5&  \\
 19&  &13~36~57.3~& -29~53~39.6~&     469.5\pm23.3    & 161.1\pm14.1 & 195.5\pm15.5 &  113.0\pm12.5 &  \\
 20&  &13~36~57.5~& -29~47~28.3~&      76.8\pm14.0    &  &  30.4\pm09.0&  29.3\pm09.3&   \\

 21&  &13~36~57.9~& -29~53~02.9~&      75.9\pm10.6    & 60.2\pm09.3 &  &  &   \\
 22&  &13~36~57.9~& -29~49~23.2~&      35.2\pm08.3    &  &  &   &  \\
 23&  &13~36~58.2~& -29~51~24.5~&      25.3\pm06.8    &  &  &   &  \\
 24&  &13~36~58.2~& -29~52~15.8~&     169.5\pm16.3    & 97.7\pm12.7 &  57.0\pm10.5&   &  \\
 25&  &13~36~58.3~& -29~48~33.0~&     106.4\pm14.7    &  &  48.4\pm10.1& 34.0\pm9.2  &  \\

 26&  &13~36~58.4~& -29~51~04.8~&     112.8\pm12.5    &  &  50.5\pm08.6 & 55.9\pm09.2 &  \\
 27&  &13~36~59.1~& -29~53~36.4~&      22.2\pm06.4    & 21.1\pm06.3 &  &  &   \\
 28&  &13~36~59.5~& -29~49~58.9~&    1085.2\pm35.4    & 357.3\pm21.1 & 427.5\pm22.4  & 300.4\pm19.3 & {\rm H17}  \\
 29&  &13~36~59.5~& -29~52~04.0~&      46.6\pm12.2    & 43.9\pm10.9 &  &  &  
(^{\mathrm{b}})\\
 30&  &13~36~59.8~& -29~52~05.6~&     268.7\pm21.0    &  93.4\pm14.1 & 124.3\pm14.0 & 54.7\pm09.9 &  
(^{\mathrm{b}})\\

 31&  &13~37~00.0~& -29~51~50.3~&     169.3\pm22.5    & 60.9\pm15.7 & 70.8\pm14.6 & 47.7\pm09.1 &  
(^{\mathrm{b}})\\
 32&  &13~37~00.1~& -29~53~29.9~&      29.1\pm07.1    &   &   &  & \\
 33&  &13~37~00.2~& -29~52~06.9~&      62.3\pm18.1    & 63.1\pm16.5 &    &  & 
(^{\mathrm{b}})\\
 34&  &13~37~00.3~& -29~52~05.4~&      60.0\pm20.0    & 52.0\pm19.2 &   &  & 
(^{\mathrm{b}})\\
 35&  &13~37~00.4~& -29~51~59.3~&     325.4\pm37.8    & 263.7\pm36.2 & 81.1\pm19.6  &  & (^{\mathrm{b}})\\

 36&  &13~37~00.4~& -29~50~54.0~&      68.8\pm10.9    & 70.3\pm10.8 &  &  &   \\
 37&  &13~37~00.5~& -29~51~56.3~&     272.6\pm39.0   & 196.7\pm31.7 & 73.0\pm23.0  &  & 
(^{\mathrm{b}})\\
 38&  &13~37~00.6~& -29~51~46.8~&      57.2\pm12.9    &  & 33.6\pm09.9  &  & 
(^{\mathrm{b}})\\
 39&  &13~37~00.7~& -29~53~19.8~&     309.0\pm19.8    & 106.9\pm11.9 & 122.8\pm13.0  & 79.3\pm10.7 &  \\
 40&  &13~37~00.7~& -29~51~59.3~&     181.1\pm39.0   &  158.4\pm33.3 &   &  & 
(^{\mathrm{b}})\\

 41&  &13~37~00.7~& -29~52~04.1~&     111.7\pm32.0   & 90.5\pm27.5 &   &  & 
(^{\mathrm{b}})\\
 42&  &13~37~00.7~& -29~52~06.0~&     168.6\pm28.9   &  & 84.0\pm16.3  & 60.0\pm09.8 & 
(^{\mathrm{b}})\\
 43&  &13~37~00.9~& -29~51~55.8~&     689.9\pm40.1
   &  201.2\pm28.2& 264.9\pm24.0  & 238.1\pm18.4 & {\rm IR~nuclear~peak} (^{\mathrm{b}})\\
 44&  &13~37~01.0~& -29~52~02.7~&    1215.2\pm44.9   &  556.0\pm32.9& 514.8\pm28.5  & 144.5\pm14.2 & 
(^{\mathrm{b}})\\
 45&  &13~37~01.1~& -29~52~45.8~&     80.5\pm11.0    &  41.0\pm08.0& 23.9\pm06.8  & 18.8\pm05.7 & 
{\rm H21}  \\

 46&  &13~37~01.1~& -29~51~51.9~&      64.7\pm15.8    &  & 53.4\pm11.0  &  & 
(^{\mathrm{b}})\\
 47&  &13~37~01.2~& -29~54~49.5~&     108.8\pm12.7    &  106.7\pm12.6&  &  &   \\
 48&  &13~37~01.3~& -29~52~01.9~&      42.9\pm09.5    &   &   & 35.3\pm07.7 & 
(^{\mathrm{b}})\\
 49&  &13~37~01.3~& -29~52~00.0~&      70.1\pm12.8    &  39.8\pm10.2& 26.9\pm08.1  &  & 
(^{\mathrm{b}})\\
 50&  &13~37~01.3~& -29~51~36.7~&      60.8\pm10.5    &   &  & 53.0\pm09.7 &  \\

   \noalign{\smallskip}
            \hline
         \end{array}
     $$
\begin{list}{}{}
\item[$^{\mathrm{a}}$] Exposure time = 49.497 ks
\item[$^{\mathrm{b}}$] Contributing to the nuclear source H19, unresolved by {\it ROSAT}
\end{list}
   \end{table*}

  \begin{table*}
         \label{}
   $$
         \begin{array}[width=textwidth]{rcccrrrrl}
            \hline
            \noalign{\smallskip}
           {\mathrm{No.}} & &
          {\mathrm{R.A.(2000)}}     & {\mathrm{Dec.(2000)}}
                 & {\mathrm{net~counts}}^{\mathrm{a}}
                 & {\mathrm{0.3-1~keV}}
                 & {\mathrm{1-2~keV}}
                 & {\mathrm{2-8~keV}}
                 & {\mathrm{notes}} \\
            \noalign{\smallskip}
            \hline
            \noalign{\smallskip}
 51&  &13~37~01.5~& -29~53~26.6~&     379.3\pm21.8    &  56.8\pm09.6& 244.6\pm17.6  & 77.7\pm10.5  & {\rm H20}  \\
 52&  &13~37~01.6~& -29~51~28.2~&     511.1\pm25.9    &  216.4\pm17.4& 199.1\pm16.6  &100.1\pm12.0  & {\rm H23}  \\
 53&  &13~37~01.7~& -29~47~42.4~&     206.5\pm19.1    &  22.7\pm07.5 & 114.5\pm13.8 & 70.3\pm12.5 &  \\
 54&  &13~37~01.8~& -29~51~13.0~&      24.3\pm07.3    &  24.7\pm07.0&   &   & \\
 55&  &13~37~02.0~& -29~55~18.2~&     319.7\pm20.0    &  103.3\pm12.3& 99.6\pm11.4  & 117.0\pm12.9 & {\rm H22}  \\

 56&  &13~37~02.2~& -29~55~06.2~&      40.7\pm08.7    &  & 19.3\pm6.0 &   &  \\
 57&  &13~37~02.2~& -29~49~52.3~&      28.8\pm07.5    &  &   &  &  \\
 58&  &13~37~02.4~& -29~51~26.2~&      75.6\pm10.4    &  56.9\pm09.4& 17.8\pm05.8  &  &  \\
 59&  &13~37~02.5~& -29~53~19.3~&      80.7\pm10.5    &  & 42.3\pm08.0 & 24.2\pm06.4  &  \\
 60&  &13~37~03.3~& -29~52~26.9~&      45.9\pm08.6    &  & 22.3\pm06.2 &   &  \\

 61&  &13~37~03.9~& -29~49~30.2~&     300.9\pm19.8    &  109.4\pm12.3 &121.3\pm13.2  & 70.4\pm10.4 &  \\
 62&  &13~37~04.3~& -29~54~03.8~&    1094.4\pm36.1    &  330.6\pm20.1& 555.1\pm26.1  &208.4\pm16.5  & {\rm H26}  \\
 63&  &13~37~04.4~& -29~51~30.5~&      43.2\pm08.2    &  20.8\pm06.1 &   &   &  \\
 64&  &13~37~04.4~& -29~51~21.5~&    1431.1\pm39.9    &  441.3\pm23.1& 718.7\pm28.5  & 270.9\pm17.9 & {\rm H27a} \\
 65&  &13~37~04.7~& -29~51~20.4~&      67.5\pm10.7    &  23.5\pm06.7& 26.5\pm07.8  & 20.6\pm06.0 & {\rm H27b} \\

 66&  &13~37~04.8~& -29~48~51.1~&      29.5\pm08.2    &  &  &  &   \\
 67&  &13~37~05.5~& -29~52~34.1~&      76.9\pm10.2    &  & 32.4\pm07.1  & 44.1\pm08.0  & \\
 68&  &13~37~06.1~& -29~55~14.7~&      35.9\pm08.3    &  27.6\pm07.3&   &  &  \\
 69&  &13~37~06.2~& -29~54~44.4~&      27.1\pm07.1    &  22.3\pm06.4&   &  &  \\
 70&  &13~37~06.2~& -29~52~32.1~&      89.5\pm11.2    &  87.8\pm11.1&   &  &  \\

 71&  &13~37~07.1~& -29~51~01.6~&     679.6\pm29.1    &  24.4\pm07.1& 286.5\pm19.6  & 366.8\pm21.4 &  \\
 72&  &13~37~11.9~& -29~52~15.6~&      29.1\pm07.3    &  24.0\pm06.7&   &  &  \\
 73&  &13~37~12.5~& -29~51~40.0~&      30.8\pm07.2    &  &  &   &  \\
 74&  &13~37~12.6~& -29~51~55.0~&     105.6\pm12.6    &  & 41.8\pm08.4  & 61.0\pm09.9 &  \\
 75&  &13~37~12.9~& -29~50~12.2~&      36.5\pm08.1    &  31.6\pm07.7&   &  &  \\

 76&  &13~37~14.5~& -29~51~49.0~&      95.3\pm11.8    &   & 54.1\pm09.1 & 25.7\pm07.1 &  \\
 77&  &13~37~14.7~& -29~49~44.2~&      53.6\pm09.0    &   & 25.2\pm06.6 &  &  \\
 78&  &13~37~16.4~& -29~49~39.1~&     397.9\pm23.9    &  95.1\pm12.1& 166.6\pm15.9  & 134.8\pm14.6 & {\rm H29}  \\
 79&  &13~37~17.2~& -29~51~53.5~&      87.3\pm11.0    &  59.1\pm09.3 & 25.4\pm06.9 &  &  \\
 80&  &13~37~18.0~& -29~52~11.8~&      37.9\pm08.1    &  21.8\pm06.2 &  &  &  \\
 81&  &13~37~19.7~& -29~51~31.7~&      66.4\pm10.5    &   & 19.0\pm05.9 & 45.9\pm08.9 &  \\
   \noalign{\smallskip}
            \hline
         \end{array}
     $$
\begin{list}{}{}
\item[$^{\mathrm{a}}$] Exposure time = 49.497 ks
\end{list}
   \end{table*}

\end{document}